\documentclass[sigconf]{acmart}

\AtBeginDocument{%
  }

\setcopyright{acmlicensed}
\copyrightyear{2025}
\acmYear{2025}
\acmDOI{XXXXXXX.XXXXXXX}

\usepackage{amsmath}
\usepackage{graphicx}
\usepackage{booktabs}
\usepackage{subcaption}
\usepackage{multirow}
\usepackage{svg}
\usepackage[table]{xcolor}
\usepackage{makecell} 
\usepackage{fancybox,framed} 

\definecolor{novicegreen}{RGB}{255, 201, 215}   
\definecolor{interblue}{RGB}{200,220,255}     

\newcommand{\badge}[2]{%
  \colorbox{#1}{\strut #2}%
}
\newcommand{\Formative}[1]{\badge{interblue}{#1}}   
\newcommand{\toolname}{\textsc{ConversAR}}

\begin{document}

\title{Practicing a Second Language Without Fear: Mixed Reality Agents for Interactive Group Conversation}

\author{Mariana Fern\'andez-Espinosa}
\email{mferna23@nd.edu}
\orcid{0009-0004-1116-2002}
\affiliation{%
  \institution{University of Notre Dame}
  \city{Notre Dame}
  \state{IN}
  \country{USA}
}

\author{Kai Zhang}
\authornotemark[1]
\email{kzhang23@nd.edu}
\affiliation{%
  \institution{University of Notre Dame}
  \city{Notre Dame}
  \state{IN}
  \country{USA}
}

\author{Jad Bendarkawi}
\authornote{indicates equal contribution by authors}
\email{jadb@alumni.princeton.edu}
\affiliation{%
  \institution{Princeton University}
  \city{Princeton}
  \state{New Jersey}
  \country{USA}
}

\author{Ashley Ponce}
\authornotemark[1]
\email{ap36@princeton.edu}
\affiliation{%
  \institution{Princeton University}
  \city{Princeton}
  \state{New Jersey}
  \country{USA}
}
\author{Sean Chidozie Mata}
\email{sm5607@alumni.princeton.edu}
\affiliation{%
  \institution{Princeton University}
  \city{Princeton}
  \state{New Jersey}
  \country{USA}
}
\author{Aminah Aliu}
\email{aa1237@alumni.princeton.edu}
\affiliation{%
  \institution{Princeton University}
  \city{Princeton}
  \state{New Jersey}
  \country{USA}
}

\author{Lei Zhang}
\email{lei.zhang@njit.edu}
\affiliation{%
  \institution{New Jersey Institute of Technology}
  \city{Newark}
  \state{New Jersey}
  \country{USA}
}

\author{Francisco Fern\'andez Medina}
\email{ffernanm@uc.cl}
\affiliation{%
  \institution{P. Universidad Catolica de Chile}
  \city{Santiago}
  \country{Chile}
}

\author{Elena Mangione-Lora}
\email{mangione.5@nd.edu}
\affiliation{%
  \institution{University of Notre Dame}
  \city{Notre Dame}
  \state{IN}
  \country{USA}
}

\author{Andrés Monroy-Hernández}
\email{andresmh@princeton.edu}
\affiliation{%
  \institution{Princeton University}
  \city{Princeton}
  \state{New Jersey}
  \country{USA}
}

\author{Diego G\'omez-Zar\'a}
\orcid{0000-0002-4609-6293}
\email{dgomezara@nd.edu}
\affiliation{%
  \institution{University of Notre Dame}
  \city{Notre Dame}
  \state{IN}
  \country{USA}
}

\renewcommand{\shortauthors}{}

\begin{abstract}
Developing speaking proficiency in a second language can be cognitively demanding and emotionally taxing, often triggering fear of making mistakes or being excluded from larger groups. While current learning tools show promise for speaking practice, most focus on dyadic, scripted scenarios, limiting opportunities for dynamic group interactions. To address this gap, we present \toolname{}, a Mixed Reality system that leverages Generative AI and XR to support situated and personalized group conversations. It integrates embodied AI agents, scene recognition, and generative 3D props anchored to real-world surroundings. Based on a formative study with experts in language acquisition, we developed and tested this system with a user study with 21 second-language learners. Results indicate that the system enhanced learner engagement, increased willingness to communicate, and offered a safe space for speaking. We discuss the implications for integrating Generative AI and XR into the design of future language learning applications.
\end{abstract}

\ccsdesc[500]{Applied computing~Interactive learning environments}
\ccsdesc[500]{Human-centered computing~Mixed / augmented reality}

\keywords{Mixed Reality (MR), Extended Reality (XR), Embodied
Agents, Second Language Acquisition, Language Learning}



\maketitle

\section{Introduction}
\label{introduction}
In an increasingly globalized society, second language acquisition (SLA) has become indispensable, not just for academic credentials or career growth, but as a tool for integration, cross‐cultural understanding, and navigating migratory transitions \cite{pokrovskaya2022academic, seven2020motivation, esser2006migration, kuo2006linguistics, crabbe2003quality, ahtif2022role}.
SLA is a cognitive learning process \cite{dekeyser2007practice, zhang2024exploring} where developing different skills determines language proficiency \cite{lightbown2021languages, selinker2008second}. Within these skills, speaking is one of the most challenging competencies for learners \cite{horwitz1986foreign, macintyre1994subtle}. Unlike reading or writing, which allow time for reflection and revision, speaking requires real-time language production, immediate comprehension, and dynamic interaction with conversational partners \cite{zhang2009role,krashen1982principles}.

\begin{figure*}
\centering
\includegraphics[width=0.9\textwidth]{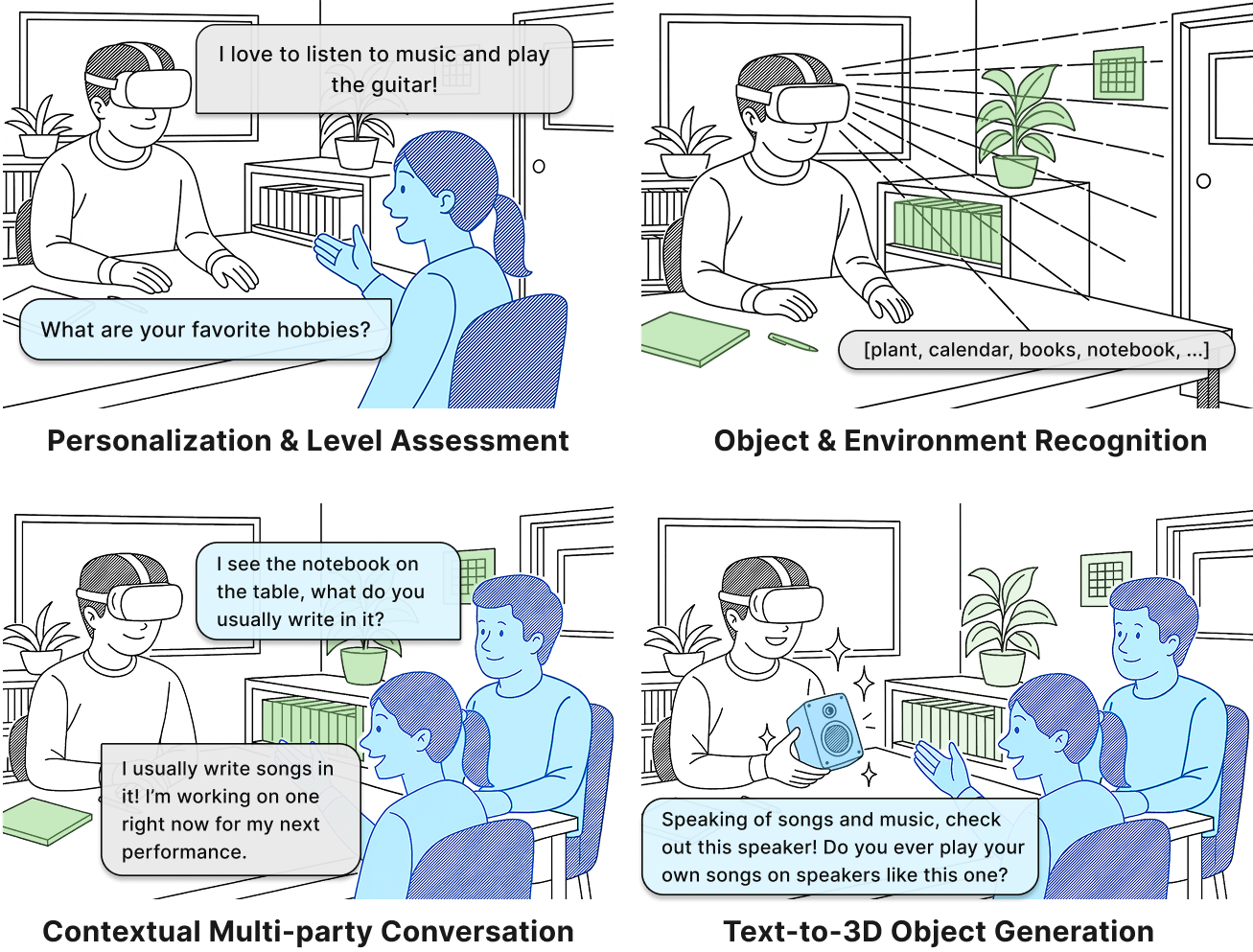}
  \label{fig:teaser}
  \caption{\textbf{\toolname{}} enables second language learners to engage in group conversations with embodied AI agents tailored to their proficiency level and personal interests. The system provides real-time corrective feedback and grounds conversations in the learner’s physical environment by recognizing real-world objects (e.g., plant, calendar, notebook). It also dynamically generates 3D digital props (e.g., speaker) informed by realia-based pedagogical theory. These tangible, contextual objects serve as conversational triggers that foster deeper oral expression, sustained interaction, and meaningful language use.}
  \Description{[Infographic of System Use]
  A four-panel illustration showing how XR language learning systems can support conversation. 
  Top-left: a learner wearing a VR headset talks to a virtual agent, who asks about hobbies (Personalization & Level Assessment). 
  Top-right: the system detects objects in the room such as a plant, books, and a notebook (Object & Environment Recognition). 
  Bottom-left: two agents join the conversation, asking about a notebook on the table, and the learner replies they write songs in it (Contextual Multi-party Conversation). 
  Bottom-right: a virtual 3D speaker appears, and an agent asks about playing songs on it (Text-to-3D Object Generation). 
  Together, the panels illustrate how personalization, contextual awareness, multi-party dialogue, and generative content can create immersive, situated language learning experiences.
  }
\end{figure*}

The development of speaking skills is not only cognitively demanding, but also affectively influenced by factors such as fear of public exposure, anxiety, and frustration, all of which can hinder oral proficiency \cite{tavares2016role, macintyre1994subtle, kiruthiga2022impact}. In addition, opportunities to practice with native speakers, studying abroad or hiring language tutors are often scarce and financially inaccessible \cite{pinar2016second}.

As a response, Computer-Assisted Language Learning (CALL) systems have emerged to address these challenges \cite{10.1145/1500879.1500947, mirani2019review, chen2022application}. Popular tools include mobile applications such as Duolingo \cite{duolingo}, which support vocabulary and pronunciation practice, as well as social platforms that connect learners with peers \cite{tandem_app} or tutors for conversational practice \cite{xia2022rlens}. Other systems leverage Large Language Models (LLMs) to create chatbots, agent tutors, or provide corrective training for oral presentations \cite{miroyan_analyzing_2025, ouyang_effects_2024, cha_chop_2024}.

However, these systems often fail to capture the complexity and unpredictability of real-world communication due to their reliance on pre-defined scenarios and limited contextual awareness \cite{pan2025ellma, duolingo, ouyang_casemaster_2025}. Speaking in authentic contexts requires adapting to dynamic social situations, elements that scripted drills rarely reproduce. Extended Reality (XR) emerges as an avenue for more realistic and context-rich speaking scenarios, enabling learners to engage in immersive role-plays, embodied agent interactions, and simulated environments that approximate real-world settings \cite{pan2025ellma}.

Despite this potential, most of these systems remain limited in two ways. First, they primarily emphasize dyadic (one-on-one) learner–agent interactions, overlooking other forms of more complex social dynamics such as group interactions. Second, many of these systems immerse learners in imagined scenarios that often lack alignment with their real-world environments and lived experiences \cite{cantone2023contextualized}. In contrast, situated learning, grounded in the learner’s real-world context, enhances engagement and fosters adaptability \cite{lave1991situated, brown_situated_1989, anderson_situated_1996}.

Recent advances in Mixed Reality (MR) and Generative AI present an opportunity to reimagine how language learners engage in speaking practice. MR technologies can situate learners in an immersive, educational environment that seamlessly blends physical surroundings with digital content~\cite{billinghurst2002augmented}, while Generative AI enables dynamic responses and personalized feedback. Together, these technologies can move beyond pre-defined scripts to create adaptive conversational partners that can utilize rich contextual cues to support more natural communication. Yet, the potential of combining MR and AI to scaffold group language learning remains underexplored in Human-Computer Interaction research.

To address this gap, we present \toolname{}, a Mixed Reality system that leverages Generative AI and MR to support situated and personalized group conversations. \toolname{} integrates embodied AI agents, scene recognition, and generative 3D props that ground practice conversations in the learner’s physical surroundings. Drawing on formative input from experts in language acquisition, we designed \toolname{} to foster engagement, reduce fear of speaking, and expand learners’ opportunities for authentic group dialogue. Our findings indicate that practicing group conversations with \toolname{} fostered a safe environment that encouraged and motivated learners’ willingness to communicate. In addition, the presence of generative 3D props helped sustain and deepen the conversations.

Overall, our contributions are threefold:
\begin{itemize}
    \item A formative study with ten SLA educators and researchers examined the challenges L2 learners face in group speaking interactions and the limitations of current techniques addressing these challenges. 
    \item The design requirements derived from the formative study and prior research in HCI and SLA, and the implementation of \toolname{}, a Mixed Reality system that supports group conversations between an L2 learner and two AI-driven NPCs. The system adapts to the user’s speaking abilities, interests, and environmental context and leverages GenAI for dynamic object creation and interaction.  
    \item An empirical evaluation of \toolname{} with twenty-one L2 learners, complemented by expert assessments from six SLA specialists. These evaluations demonstrate the benefits and challenges of MR-based group conversations and provide design implications for future systems.  
\end{itemize}

\section{Related Work}\label{literature_review}
We draw from prior literature in Second Language Acquisition (SLA), Extended Reality (XR) for language learning, conversational systems powered by Large Language Models (LLMs), and Generative AI in the education settings.

\subsection{Second Language Acquisition as a Situated and Social Process}

Learning a second language (L2) is widely recognized as a situated, social, and interaction-driven process \cite{gass2014input}. The Interaction Hypothesis posits that conversational exchanges enhance language development by clarifying meaning, providing feedback, and guiding adjustments \cite{gass2014input}. Sociocultural Theory emphasizes that learners develop skills through mediated social activity and interaction with others \cite{lantolf2000introducing, lopez2008sociocultural}. Together, these theories suggest that interactive, conversation-based learning is more effective than traditional textbook, classroom instruction, or rote approaches.

Group conversations in classrooms promote co-construction of meaning, fluency development, and low-stakes speaking practice \cite{poehner_group_2009, ernst_talking_1994, felicity_speaking_2018}. Situated learning theory emphasizes that embedding language learning in authentic contexts bridges the gap between theoretical knowledge and practical use, facilitating transfer to real-world communication \cite{lave1991situated}. Such contexts allow learners to construct and apply language in more meaningful and engaging ways \cite{cobb1999cognitive}. Despite their pedagogical value, group conversations can be cognitively and emotionally demanding for second language learners \cite{tavares_role_2016, kim_exploring_2024} which can negatively impact learning outcomes \cite{felicity_speaking_2018}. 


To address these challenges, various L2 digital tools have emerged. Mobile apps (e.g., Duolingo, Babbel) use gamification and spaced repetition to reinforce vocabulary and grammar \cite{duolingo, 2018L2P}. Platforms like Tandem enable informal peer or native-speaker exchanges \cite{nushi_tandem_2020}, while AI chatbots provide one-on-one conversational practice with corrective feedback \cite{pan2025ellma}. However, these tools often lack contextual awareness, limiting sustained engagement and opportunities for embodied, multimodal interactions that promote more natural communication.

In response, a growing body of research in HCI and Second Language Acquisition (SLA) has identified the intersection of Extended Reality (XR) and Artificial Intelligence (AI) as an avenue for creating realistic and engaging language learning environments \cite{schnitzer2025language, hirzle2023xr, divekar2022foreign}.

\subsection{Extended Reality for Learning}

Extended Reality (XR) enhances learning by supporting embodied interaction, contextual immersion, and multimodal engagement. In education, it has been shown to improve memory retention \cite{lukianova2025picture}, provide high-fidelity simulations for technical skills, foster cultural competency, and clarify abstract concepts through interactive 3D representations, making it a powerful tool for experiential and situated learning.

A related approach is the use of realia, physical objects that create multi-sensory, context-rich experiences \cite{bably2017using}. Realia links abstract language to tangible referents, enhancing comprehension, retention, creativity, and motivation \cite{hadi2018effectiveness}, while exposing learners to culturally relevant artifacts \cite{bala2015positive}. Yet, it faces limitations, including low personalization, restricted interactivity, and limited alignment with learners’ conversational needs \cite{lee2021review}. XR can address these gaps by providing dynamic, interactive, and tailored contextualization, extending the benefits of realia into immersive digital environments.

\subsection{Extended Reality for SLA}
In modern SLA approaches, XR has been applied to create engaging, context-rich learning experiences, including role-playing in simulated scenarios, situated self-talk \cite{hollingworth2023fluencyar}, interaction with 3D objects for vocabulary reinforcement \cite{schnitzer2025language}, gamified activities, embodied tutoring \cite{dai2024designing}, and NPC-guided conversations.

Many XR systems in SLA have primarily focused on vocabulary acquisition. For example, Spelland \cite{hsu2023spelland} uses MR to spell nearby objects, enhancing engagement and retention, while ARLang \cite{caetano2023arlang} overlays visual annotations onto physical objects for contextualized learning. XR also supports speaking and pronunciation; embodied conversational agents in Social Virtual Reality (VR) guide interactive role-plays, offering realistic, low-stakes fluency practice \cite{pan2025ellma, qiu2023trends}.

Although XR often emphasizes individual or dyadic interactions, recent work explores collaborative learning. VR systems simulate real-world scenarios (e.g., restaurants) for group tasks that reinforce vocabulary, discussion, and role-playing \cite{cantone2023contextualized, adjagbodjou2025s}. CILLE enables multi-learner interactions with multiple embodied agents, supporting vocabulary, cultural understanding, and speaking skills \cite{divekar2022foreign}. Few studies, however, examine how multiple agents can coordinate with humans to simulate group conversation dynamics in contextualized environments \cite{bendarkawi2025conversar}.

\subsection{Large Language Models and Conversational Agents in SLA}
Recent studies have leveraged LLMs to enhance CALL systems by providing real-time feedback, grammar correction, fluency and pronunciation practice, and emotional scaffolding to sustain learner motivation \cite{li2024systematic, han2024chatgpt, kim_exploring_2024}. LLM-powered chatbots, often embodied as Non-Player Characters (NPCs) in game-based or simulated environments, enable dyadic learner–NPC interactions that simulate face-to-face communication and create immersive practice settings \cite{karaosmanoglu2024language, zhao2024language, pan2025ellma}.

For example, The Language of Zelda integrates NPCs for French practice through gameplay and quests \cite{karaosmanoglu2024language}, while Language Urban Odyssey provides corrective feedback within a culturally rich fictional city \cite{zhao2024language}. Also, collaborative gamification approaches, as used in Crystallize \cite{culbertson2016crystallize} and LingoLand \cite{seow2023lingoland}, immerses learners in shared tasks or personalized real-world scenarios, promoting speech production.

Despite these engaging experiences, such systems are limited in capturing learners’ real-world contexts. Interactions occur in fictional environments disconnected from physical and social realities, restricting support for situated language learning, which relies on contextual cues, real-world references, and non-verbal signals.

\subsection{Generative AI for Second Language Acquisition}
Generative AI (GenAI) has demonstrated growing potential in education by fostering new, engaging learning experiences and enhancing traditional learning methods \cite{zhang_empowering_2025}. Its ability to produce multimodal content such as text, images, 3D models, and videos enables more interactive, adaptive, and engaging learning environments that support the learning development of the students \cite{zhang_empowering_2025, morita_genaireading_2025, liu_singakids_2025, zhuang_enhancing_2025}. 

In educational settings, Liu et al. \cite{liu2025bricksmart} developed `BrickSmart,' a system that leverages GenAI to teach children spatial language vocabulary. The system combines `text-to-3D' generation with interactive, guided block-building activities to scaffold learning through multimodal instruction. GenAI has also been used to enhance both instructional materials and learner engagement \cite{zhang_empowering_2025}. For instance, Simonsen et al. \cite{bedi2023using} leveraged genAI for creating image–text narratives for Icelandic L2 learners, creating pedagogical, multimodal materials aligned with curricular goals. Similarly, generative `text-to-video' tools have been employed to assist learners in producing multimedia presentations of learning content with the purpose of enhancing writing and listening skills \cite{takeda2024enhancing}.

Nevertheless, the use of GenAI to support group conversations in SLA remains largely underexplored. Our approach aims to examine how aligning with the pedagogical strategy of realia can stimulate contextualized interaction and enhance the language experience.

\section{Formative Study}
To inform the design of \toolname{}, we conducted a formative study with a group of professors and instructors of Second Language Acquisition (SLA). Through semi-structured interviews, we explored the main challenges in students' speaking development in group conversations, effective teaching techniques, and common challenges they face with digital tools for SLA. The study was reviewed and approved by \anon{the University of Notre Dame}’s Institutional Review Board (Protocol \anon{25-06-9350}) These insights helped us ground our design decisions in established learning goals and practices.

\subsection{Method and Participants}
\subsubsection{Recruitment}
We recruited 10 educators through a combination of personal outreach, emails to academic institutions in the United States and Mexico, and posts on social media channels. Participants were selected based on a minimum of two years of teaching experience. One of the authors conducted 45-60 minute interviews with each participant on Zoom. Each participant received a \$25 gift card as compensation for their time.

\subsubsection{Participants}
As shown in Table \ref{tab:educator_profiles}, participants included SLA instructors and researchers from academic institutions. Their teaching experience ranged from 3 to 35 years. They had taught Spanish, English, Chinese, Italian, and Korean across beginner (A1, A2), intermediate (B1, B2), and advanced proficiency levels (C1-C2), as defined by the Common European Framework of Reference for Languages (CEFR) \cite{council2001common}. 

\begin{table}[!htb]
\centering
\caption{Profiles of second language educators interviewed}
\label{tab:educator_profiles}
\resizebox{\columnwidth}{!}{%
\begin{tabular}{
  p{0.04\linewidth}  
  p{0.09\linewidth}  
  p{0.21\linewidth}  
  p{0.20\linewidth}  
  p{0.13\linewidth}  
  p{0.17\linewidth}  
  p{0.12\linewidth}  
}
\toprule
\textbf{\#} &
\textbf{Gender} &
\makecell[l]{\textbf{Job Title}} &
\makecell[l]{\textbf{Work}\\\textbf{Affiliation}} &
\makecell[l]{\textbf{Years of}\\\textbf{Experience}} &
\makecell[l]{\textbf{Target}\\\textbf{Language}} &
\makecell[l]{\textbf{Instructional}\\\textbf{Level}} \\
\midrule
01 & Female & Teacher & University / Private Classes & 25 years & English, Spanish & A2, B1, B2 \\
02 & Male & Teacher (PhD Candidate Spanish) & University & 3--5 years & Spanish & A1, A2, B1, B2, C1, C2 \\
03 & Female & Teacher (PhD Candidate Spanish) & University & 3--5 years & English, Spanish, French, Italian & A1, A2, B1 \\
04 & Female & Professor Researcher & University & 35 years & Spanish & B1, B2 \\
05 & Female  & Teacher & University & 3--5 years & English & B1, B2, C1 \\
06 & Female  & Professor & Community College & 30  years & English & A1-C2 \\
07 & Female & Professor & University & Over 10 year & English & B1-C2 \\
08 & Female & Professor & University & Over 10 year & Chinese and Korean & A1-A2 \\
09 & Female & Professor & University & Over 10 year & Italian & A1-A2 \\
10 & Female & Professor & University & Over 10 year & Spanish & A1-A2 \\
\bottomrule
\end{tabular}%
}
\label{tab:educator_profiles}
\end{table} 

\subsubsection{Procedure}
The interviews followed a semi-structured script with four main stages. First, participants discussed the speaking challenges they observed among their students when learning a second language and participating in group conversations. Second, we explored the limitations of the digital tools they currently use to support such interactions. Next, participants reflected on these challenges in greater depth, described strategies they use to address them, and identified gaps in their experiences. In the final section, we provide a walkthrough of an initial prototype of \toolname{}, showing how students would chat with two virtual agents in a Mixed Reality environment. We explained that the design space was entirely open, with the system and its agents free to adopt any roles, behaviors, or interaction styles. We encouraged participants to brainstorm ideas without concern for technical limitations. Participants proposed a range of features and capabilities to best support language learners. Each interview concluded with a discussion of any concerns or ethical considerations regarding the integration of such systems into their teaching practice.

\subsubsection{Data Analysis}
After transcribing the interviews \cite{aws2024transcribe}, three authors conducted a thematic analysis \cite{braun2006using} to develop an initial set of codes. These codes were refined and consolidated into a codebook through affinity mapping sessions conducted online using a Miro board \cite{nielsen2024affinity, miro2025}. Using this codebook, all interviews were coded, and through iterative analysis, four main themes emerged.

\subsection{Key Insights}
Our formative study revealed that learning a new language, particularly the speaking component, is not merely a cognitive task but also a deeply emotional and physiological process. The interviews identified a range of affective barriers that hinder students' speaking development, including fear, frustration, embarrassment to participate in group conversation, and lack of confidence and motivation. Additionally, participants identified several key challenges when employing current digital tools. 

\subsubsection{\textbf{KI1}: Students fear mistakes and the judgment of others when speaking}
A prominent challenge identified by participants was the persistent fear among students of making mistakes during speaking activities. They observed that this fear inhibits students' confidence, particularly in group settings. As one instructor explained (\Formative{P05}), \textit{``students are always scared to make mistakes, simply because they are afraid to say it wrong.''} This fear is amplified in classroom environments where learners feel they are under constant evaluation by their professors. Another participant noted that many students (\Formative{P01}) \textit{``are always scared to speak''} because they worry about \textit{``...repercussions on their grades'',} requiring the instructor to clarify that class conversations are not graded.

\subsubsection{\textbf{KI2:} Students fear being left out when working with peers of higher proficiency.}
Participants observed that students with low self-confidence often remain silent during group conversations, especially when they perceive other speakers as more proficient. In these situations, more advanced speakers tend to dominate the discussion and rarely invite more silent peers to contribute. One educator (\Formative{P10})  explained that \textit{``if they feel the others are better, they just step back and stay quiet''}, noting that the lack of direct invitations to speak often leaves less confident learners on the margins of the conversation. Over time, this dynamic reinforces hesitation and limits opportunities for practice. Another participant (\Formative{P06})  reflected, \textit{``They do not interrupt, and if nobody asks for their opinion, they simply don't participate''}. Such patterns can create a cycle of exclusion, where struggling speakers become increasingly passive and disengaged from collaborative dialogue.

\subsubsection{\textbf{KI3:} Students fear repeating mistakes because feedback is missing}
Several participants noted that group conversations often do not provide the corrective feedback learners require to improve their proficiency. In many cases, peers do not point out errors because they do not notice them, are unsure how to correct them, or avoid mentioning them to prevent embarrassment. As one educator (\Formative{P08}) remarked, \textit{``They just let them pass, and the student keeps saying it wrong.''} Without timely correction, these mistakes can become habitual, a phenomenon participants associated with the ``fossilization of errors'' \cite{valette1991proficiency}. Another participant (\Formative{P02}) explained, \textit{``If nobody corrects them, they will keep making the same mistake over and over.''} This absence of feedback, while often unintentional, can diminish the potential benefits of group interaction by allowing inaccurate forms to persist and become fixed in learners' speech.

\subsubsection{\textbf{KI4:} Students disengage when activities feel irrelevant to their lives}
Participants noted that many students struggled to engage with classroom activities built around static materials, such as textbook dialogues or fixed role-play scenarios, which often failed to align with their interests or reflect their lived realities. Learners frequently found it difficult to visualize or relate to these imagined contexts, limiting their ability to participate meaningfully. As one educator explained, (\Formative{P04}) \textit{``... if it's not part of their world, they don’t know how to react, and the conversation goes nowhere.''} The lack of personal relevance in group conversations was a key factor in this disengagement. Participants observed that when the topics or scenarios did not connect with students' everyday experiences or personal interests, even well-designed activities failed to generate authentic dialogue. As a result, these strategies frequently fell short of fostering participation.

These key insights illustrate how students' fears and frustrations shape their experiences when learning a second language in group conversations. Together, these insights highlight the importance of designing learning environments that not only foster students' participation but also provide timely feedback, build confidence across different proficiency levels, and connect meaningfully to students' lived experiences.
\section{The \toolname{} System}
\subsection{Design Goals}
We summarize the following key design goals derived from the findings of our formative study, which guided the final design of \toolname{}.

\subsubsection{DG1: Facilitating Confidence Through Group Conversations}
Group conversations are cognitively and affectively demanding activities for L2 learners, who often experience fear of making mistakes and require time to build speaking confidence. Interviewees emphasized that such activities should promote a supportive atmosphere where learners feel less intimidated than they typically do in classroom settings. To address this challenge, the system should incorporate multiple LLM powered NPCs to emulate realistic multi-party conversations across a variety of spontaneous topics. Practicing in this virtual, controlled yet authentic environment allows learners to gradually gain more comfort in group discussions without the social pressure of interacting with human peers. Rather than replicating hierarchical roles such as professors or examiners, the system should foster camaraderie in these conversations, creating a safe and motivating space for language practice.

\subsubsection{DG2: Delivering Supportive Corrective Feedback}
Participants emphasized that the way feedback is delivered is crucial for sustaining learners' confidence and motivation. As one (\Formative{P02}) explained, \textit{``It's about how we deliver the idea that you made a mistake.''} In speaking tasks, error correction needs to be handled with care so that students remain encouraged to participate. Researchers identify several forms of corrective feedback. One common approach is \textit{explicit correction}, in which the instructor directly provides the correct form. Another is the use of \textit{recasts}, where the instructor repeats the learner's utterance without the error, modeling the right language without explicitly pointing out the error. \textit{Clarification requests} signal misunderstanding or incorrectness, prompting the learner to repeat or reformulate. Finally, \textit{metalinguistic feedback} provides comments or questions related to correctness without directly giving the answer. To address these needs, \toolname{} should implement such feedback strategies with a supportive tone. For example, when a grammar or pronunciation error occurs, the system can rephrase or repeat the learner's utterance correctly. This implicit correction approach allows learners to notice and self-adjust without feeling embarrassed or discouraged.

\subsubsection{DG3: Creating Realistic and Contextualized Conversations}
Practicing a second language is often based on abstract that lack relevance to learners' immediate experiences, limiting both engagement and the practical transfer of skills. To address this, \toolname{} should integrate elements from the learner's physical environment into conversations, fostering immersion and promoting situated learning. By detecting and referencing physical objects, the system can create scenarios that are both personally meaningful and contextually grounded, supporting the transfer of language skills to authentic contexts. 

\subsubsection{DG4: Interactive Props to Sustain Conversation}
When teaching languages, instructors often draw on learners' personal interests or familiar topics to initiate and maintain conversations. However, creating these opportunities in the classroom can be challenging, as it frequently requires imagination, access to physical props, or extensive preparation. Without concrete visual or interactive elements, learners may struggle to visualize concepts or sustain engagement, particularly when discussing abstract or unfamiliar topics. To address this challenge, \toolname{} should be able to dynamically generate virtual objects based on the evolving context of the conversation and the learner’s expressed interests. These virtual objects can serve as conversational anchors, offering shared reference points for learners and conversational agents, while encouraging learners to observe details, describe attributes, and ask follow-up questions that sustain the dialogue. By introducing virtual objects based on the conversation (e.g., a car when discussing transportation or a fruit when learning about food), \toolname{} can ground the conversation in tangible and context-rich scenarios.

\subsubsection{DG5: Adaptive Scaffolding for Language Proficiency} 
The system should tailor conversations to match learners' proficiency levels, ensuring that content remains both understandable and motivating. As described in \textbf{\textit{KI2}}, many users will get frustrated when the conversational content exceeds their comprehension level. When the content exceeds a learner's comprehension level, it can lead to frustration and reduced confidence. This requires dynamically adapting vocabulary and sentence structure to provide appropriate scaffolding that supports users' gradual skill development. To further sustain engagement, the system should employ strategies such as circumlocution, rephrasing or describing concepts in simpler terms \cite{liskin1996circumlocution}, so that learners with low proficiency and limited vocabulary can continue participating in the conversation. By adapting in real time to each learner's level, the system can maintain confidence, encourage active participation, and promote steady growth in communicative ability.

\begin{figure*}[!hbt]
    \centering
    \includegraphics[width=\textwidth]{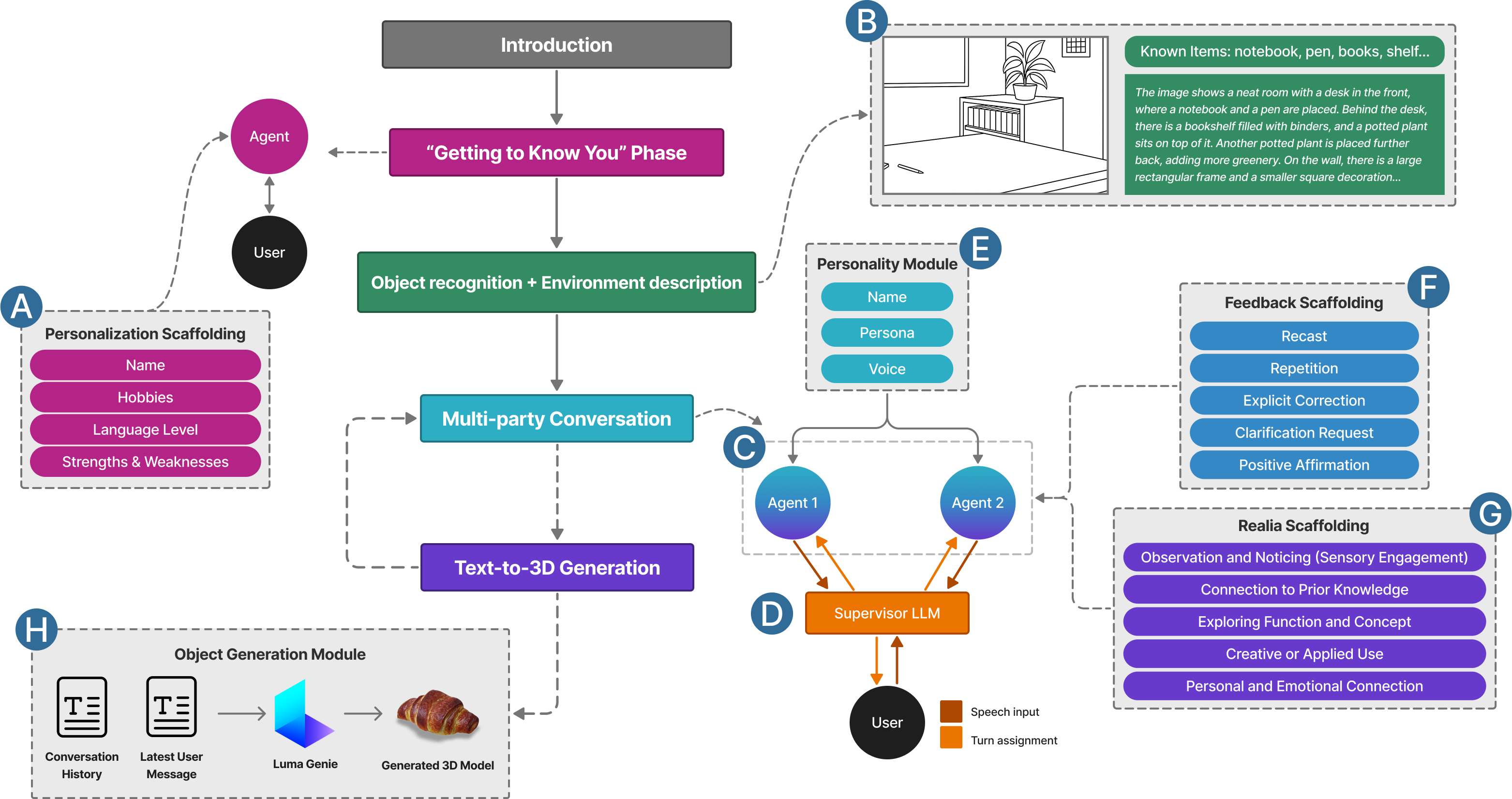}
    \caption{\textbf{\toolname{} interaction flow} illustrating how the system assesses language proficiency and interests, initiates group conversations grounded in the physical environment, generates contextual 3D props, and delivers real-time corrective feedback.}
    \label{fig:sysfeatures}
      \Description[Flowchart of a multimodal conversational learning system with labeled modules A–H.]
  {A flowchart depicts a conversational learning system integrating personalization, environment awareness, and scaffolding. The pipeline starts with Introduction to “Getting to Know You” to  Object recognition + Environment description to {Multi-party Conversation} to {Text-to-3D Generation}. 
   (A) Personalization Scaffolding: name, hobbies, language level, strengths/weaknesses. 
   (B) Environment module: visual scene shows a desk, notebook, pen, bookshelf with binders, wall frames, and potted plants; a callout lists recognized items. 
   (C–D) Multi-agent conversation: two agents converse with the user, coordinated by a Supervisor LLM; arrows indicate speech input and turn assignment. 
   (E) Personality module: name, persona, voice for agents. 
   (F) Feedback Scaffolding: recast, repetition, explicit correction, clarification request, positive affirmation. 
   (G) Realia Scaffolding: sensory engagement, connection to prior knowledge, exploring function/concept, creative/applied use, personal/emotional connection. 
   (H) Object Generation module: converts conversation history and latest user message into a generated 3D model via Luma Genie. 
   Overall, the diagram shows how personalization and multimodal input (vision + language) support interactive learning with dynamic 3D content generation.}
\end{figure*}

\subsection{Example Scenario}
In this section, we provide an example of how \toolname{} operates in a realistic context. Imagine that Alice is learning English as a second language. She’s currently at the CEFR A2 level and struggles with speaking anxiety when participating in group discussions in the target language. Wanting to practice speaking with others in a safe yet practical environment, Alice decides to use \toolname{} for a multi-party conversation session.

Alice sits down at her office desk, puts on her Meta Quest 3, opens the \toolname{} application, and selects \textit{Start}. This begins the one-on-one \textit{``Getting to Know You'' Phase}. Alice observes a single animated NPC sitting across from her (Figure \ref{fig:realsystem}a). The NPC begins the conversation by asking,
\textit{``How are you today, what's your name?''}. The NPC waits until Alice responds, pauses to think, then responds with a question about Alice's hobbies. Alice and the NPC go back and forth getting to know each other. Once 2 minutes have passed, the NPC notifies Alice that the one-on-one session has concluded and she will now move on to the \textit{Multi-party Conversation}. The system internally generates insights about Alice's language level, hobbies and interests, and strengths and weaknesses in the target language (Figure \ref{fig:sysfeatures}a) to inform the  complexity and topics of the conversation.

Alice selects the \textit{Next} button. Before the next stage is shown, the system quickly takes a snapshot of Alice's surroundings, performing object recognition and generating an environment description (Figures \ref{fig:sysfeatures}b, \ref{fig:realsystem}b). 
This information is used to enable contextually relevant responses from the NPCs.

Alice observes two animated NPCs sitting across from her (Figure \ref{fig:realsystem}c), a male NPC named Omar and a female NPC named Marta. Alice greets the NPCs and Omar responds. Noticing the headphones on the office table, Omar asks Alice about her favorite music genres. Alice is given an infinite amount of thinking time then speaks her response out loud when she is ready. Alice's response directly mentions Marta, so Marta is selected as the next NPC to speak. Alice, Marta, and Omar continue their discussion about music for a few turns, each speaking one at a time, with turns being dynamically assigned by a \textit{Supervisor LLM} based on the natural flow of dialogue.

As the conversation history has been centered around music, a realistic 3D rendered speaker appears on the table (Figure \ref{fig:realsystem}d). Omar encourages Alice to pick up the speaker and creatively reflect on what features she might add to the object to make it more useful or fun. While formulating her reply, she pinches the edges of the object with her index finger and thumb, bringing it closer for further inspection to spark new ideas (Figure \ref{fig:realsystem}e). Alice finally answers, but makes a few tense-related grammatical errors in her response. Omar provides implicit feedback by smoothly including the proper tense usage in his response and following with a positive affirmation.

Alice, Marta, and Omar continue conversing, covering various topics of interest to Alice, making references to the environmental context, and generating relevant 3D objects to drive the discussion forward. Alice continues to receive in-the-moment explicit or implicit corrective feedback when appropriate throughout. Finally, after \(\sim \)15 minutes of use, Alice says her goodbyes to Marta and Omar and closes the application, ending the \textit{Multi-party Conversation} stage and concluding the \toolname{} experience.

\subsection{Key Features}
This section introduces the key features of \toolname{}, as illustrated in the example scenario and Figure \ref{fig:sysfeatures}. These features include group conversations with LLM-based NPCs providing supportive corrective feedback, scene recognition, contextual generative AI 3D objects, and adaptation to learner preferences and linguistic level .

\subsubsection{Group Conversations using LLM-based NPCs with Supportive Corrective Feedback}
As described in \textit{DG1} and \textit{DG2}, \toolname{} facilitates realistic multi-party group conversations between the learner and two NPCs driven by ``speech-to-text'' (STT) and ``text-to-speech'' (TTS) capabilities. Designed to act as peer participants rather than instructors, these NPCs foster a casual, non-intimidating environment where learners can practice speaking without the social pressure often present in in-person environments. A \textit{Supervisor LLM} manages turn-taking to ensure balanced engagement among the NPCs and the user (Figure \ref{fig:sysfeatures}d). In the agents turn, the two NPCs converse with each other for up to three consecutive turns—a limit determined through pilot testing—while consistently addressing the learner with direct questions or invitations to contribute.

The corrective feedback strategies described in \textit{DG2} are embedded directly into the NPCs' conversational behavior (Figure \ref{fig:convo}). When errors occur, the NPCs apply supportive techniques such as recasts, clarification requests, or implicit reformulations, approaches shown to guide learners toward the correct form without disrupting the flow or causing embarrassment. Users can also ask questions or request repetition, further reinforcing active participation. By combining dynamic multi-party dialogue, balanced turn-taking, and sensitive real-time feedback, \toolname{} sustains engagement while promoting both confidence and linguistic growth.

\subsubsection{Scene Recognition} 
As described in \textit{DG3}, \toolname{} includes a scene recognition capability that captures and analyzes the learner's physical surroundings to ground conversations in a real-world context. When the application launches, it automatically captured the learner's initial view using the headset device's passthrough camera and processes this visual input through an LLM to generate a concise description of the scene and identify relevant objects within the environment.


\subsubsection{Contextual Generative AI 3D Objects}
To support DG4, \toolname{} integrates a real-time object generation feature that sustains and enriches conversations. Using an LLM, the system continuously analyzes the ongoing dialogue to identify its semantic context and determine a tangible object that can serve as a shared conversational anchor. Once identified, the virtual object is created via a text-to-3D generative AI module and placed into the learner’s mixed reality environment. Learners can freely scale, move, and interact with these objects, which act as visual and interactive prompts to encourage observation, description, and follow-up questions (Figure 3d).

\begin{figure*}[!hbt]
    \centering
    \includegraphics[width=\textwidth]{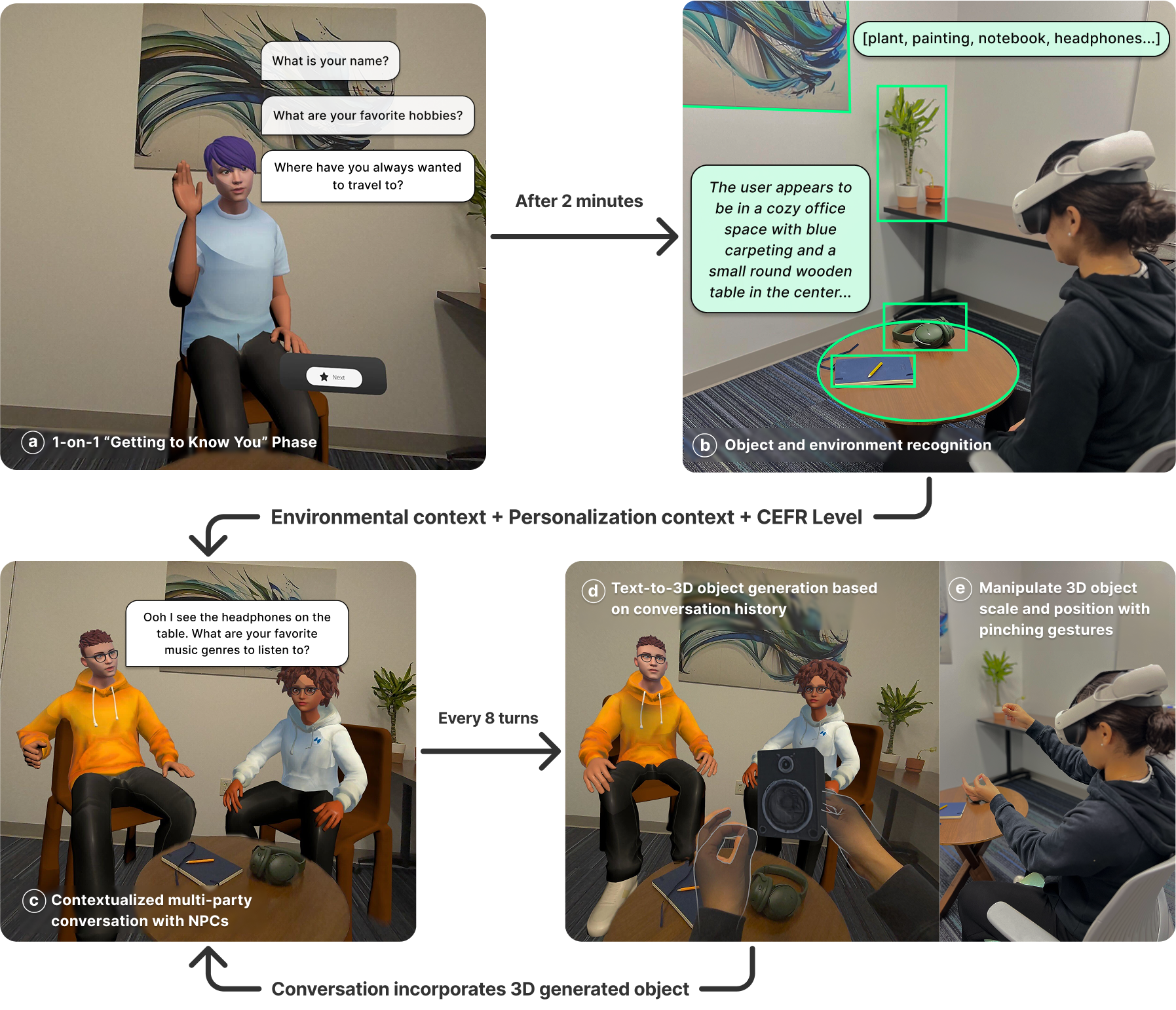}
    \caption{\textbf{System Overview of \toolname{}}. \textbf{(a)} Learners begin with a 1-on-1 warm-up conversation to assess language proficiency and personal interests. 
    \textbf{(b)} The system detects real-world objects and generates a contextual scene description. \textbf{(c)} Based on the learner’s environment and language level, an adapted group conversation with AI agents unfolds, referencing physical objects and interests. \textbf{(d, e)} \toolname{} dynamically generates 3D digital props grounded in realia pedagogical theory. These objects serve as tangible conversation anchors, fostering language development and deeper engagement.}

    \label{fig:realsystem}
    \Description{[System Overview of ConversAR]
  A multi-panel diagram illustrating an XR conversation flow for language learning. 
  (a) A learner begins a one-on-one "getting to know you" phase with an NPC. 
  (b) The system recognizes objects in the real environment (e.g., plant, painting, notebook, headphones) and integrates this context. 
  (c) Two NPCs join for a contextualized multi-party conversation, referencing objects in the space. 
  (d) Based on the dialogue, a 3D speaker object is generated and manipulated by the learner through hand gestures. 
  The pipeline shows how personalization, environmental context, CEFR level, and generated 3D objects are incorporated into conversation. 
  }
\end{figure*}

\subsubsection{Adaptation to Learner Preferences and Linguistic Level}
To ensure that conversations are both accessible and personally relevant (\textit{DG5}), \toolname{} adapts the NPCs’ speech and language to match each learner’s proficiency level and interests.. Linguistic complexity is similarly tailored, with vocabulary and sentence structures parameterized to align with the learner's speaking ability. In addition, \toolname{} incorporates a user profile generated from a series of on-boarding questions about the learner’s interests, habits, and preferred topics (Figure \ref{fig:sysfeatures}a, \ref{fig:realsystem}a). This profile is embedded into the system’s prompts, enabling NPCs to adapt not only their linguistic output but also their conversational content to align with what is most engaging and meaningful to the learner.



\subsection{Implementation}
We developed \toolname{} using Unity 6000.0.39f1 and C\#. \toolname{} ran on the Meta Quest 3 and utilized the Meta Passthrough Camera API along with the Meta XR All-in-One SDK. We implemented the NPCs using Ready Player Me \cite{readyplayerme} and animated them with Mixamo animations \cite{mixamo}.

OpenAI API was the model provider used. None of the models used were fine-tuned. In the following sections we detail the technical approaches of our implementation. Full prompts are provided in Appendix. 

\subsubsection{Object Recognition \& Scene Understanding}
We utilize the Meta Passthrough Camera API \cite{metaPCAOverview} to capture a snapshot of the user’s point of view right before the \textit{Multi-party Conversation} task. This image is passed to GPT-5 with a prompt requesting a list of prominent objects and overall environment description (Figure \ref{fig:sysfeatures}b, \ref{fig:realsystem}b). 


\subsubsection{Prompt Scaffolding}
To generate agent responses, we utilize a zero-shot prompting approach with the goal of minimizing latency \cite{sahoo_systematic_2024}. Every turn, a stand-alone base prompt (Appendix \ref{appendix:baseprompt}) takes in the full conversation history, current agent’s persona (Figure \ref{fig:sysfeatures}e), and recognized objects and scene context. We further include scaffolding blocks for personalization (Appendix \ref{appendix:profilesummarizerprompt}), corrective feedback (Appendix \ref{appendix:feedbackprompt}), and realia learning strategies (Appendix \ref{appendix:realiaprompt}) into the base prompt to align agent responses with the design goals (Figure \ref{fig:sysfeatures}a, \ref{fig:sysfeatures}f-g). The  realia scaffolding is enabled on turns in which a 3D object is generated to ensure the agent asks a rich follow-up question.



\subsubsection{STT-TTS Pipeline} We integrated the OpenAI Audio API to manage all voice-based input and output through the speech-to-text (STT) and text-to-speech (TTS) modules. For transcription, we used the gpt-4o-mini-transcribe model, and for synthesis, the gpt-4o-mini-tts model. To capture user speech, we implemented an event-based recognition system with silence detection, configured with a 2000 ms timeout and a silence threshold of 0.01f. Furthermore, users have an infinite thinking period when it is their turn, enabling them to speak at their own discretion. 

\subsubsection{Multi-party Turn Management}
We implemented a \textit{``Supervisor LLM'' } using GPT-5, which was responsible for managing turn-taking within the conversation (Figure \ref{fig:sysfeatures}d). The supervisor evaluates each dialogue exchange and determines the most appropriate next speaker among the two agents and the human participant. To guide this process, we designed a system prompt that explicitly instructed the moderator to \textit{return a singular value corresponding to the next speaker}. The supervisor followed two rules, avoiding consecutive turns by the same speaker and redirecting implicit follow-ups to the previous speaker to maintain a natural flow and balanced participation.

\subsubsection{Text-to-3D Object Generation}
We used the Genie API by Luma AI \cite{lumalabsGenie} to generate 3D objects from text prompts (Figure \ref{fig:sysfeatures}h). Using the conversation history and object generation prompt (Appendix \ref{appendix:objectsuggestionprompt}), a contextually relevant object is generated. To balance responsiveness and avoid over-generation, we triggered this process every 8 turns, based on insights from pilot sessions and testing. We use guardrails in the prompt to prevent duplicate generations and ensure the object is realistic to the context of use. Once imported into Unity, each object is attached with colliders and grab/pinch interactors from the XR Interaction Toolkit, enabling users to pick up, move, and scale objects using hand tracking.

\section{Exploratory User Study}
We conducted a controlled lab user study with 21 participants to evaluate the usability, support, and usefulness of \toolname{}. The study was reviewed and approved by \anon{the University of Notre Dame's} IRB. The study aims to answer the following questions:
\begin{itemize}
    \item[\textbf{RQ1}:] Can users successfully participate in and engage with group conversations in a second language using \toolname{}?
    \item[\textbf{RQ2}:] How useful is \toolname{} in facilitating second language speaking practice through interaction with physical and virtual objects?
    \item[\textbf{RQ3}:] What challenges do users face when using \toolname{} to practice group conversations in a second language?
\end{itemize}

\begin{figure*}[!hbt]
    \centering
\includegraphics[width=0.8\textwidth]{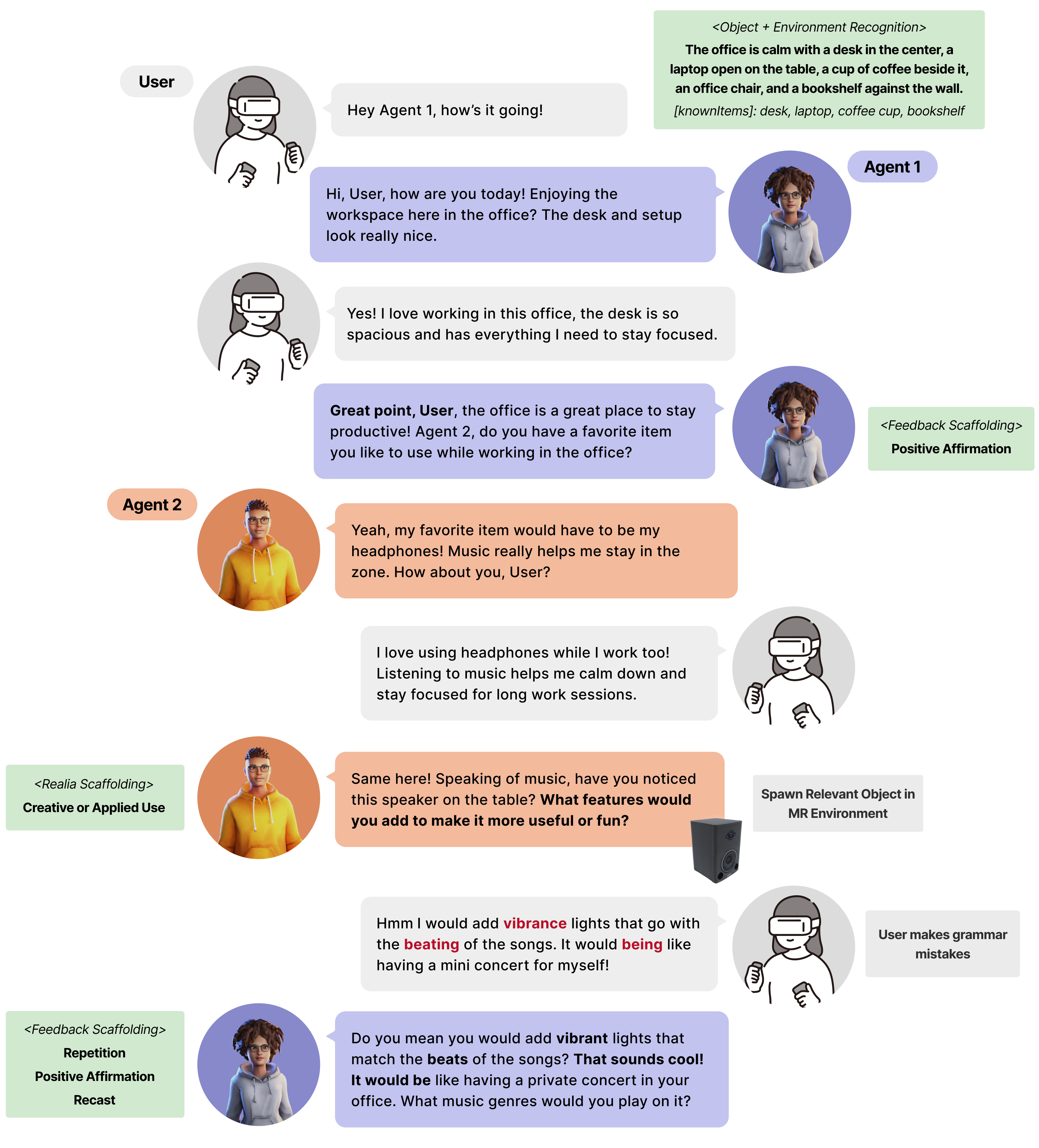}
    \caption{Example of \toolname{} conversation with corrective feedback and 3D object generation.}
    \label{fig:convo}
    \Description{[ConversAR Conversation Example]
  An example conversation in the ConversAR system where a learner interacts with two NPC agents in an office-like MR environment. 
  Agents provide positive affirmation, recasts, and scaffolding as the learner talks about headphones and music. 
  A virtual speaker object is introduced into the scene, prompting creative tasks such as imagining new features. 
  The learner makes grammar mistakes, and agents provide corrective feedback while sustaining dialogue. 
  The figure illustrates how feedback scaffolding, realia scaffolding, and environment-linked 3D object generation support situated language learning. 
  }
\end{figure*}

\subsection{Participants}
We recruited 21 participants by contacting private language schools in a Midwestern U.S. state. First, the research team contacted schools' administrators via email to explain the purpose of the research and request that they share the call for participation with their students. Additionally, we recruited participants through social media posts in groups dedicated to open conversation tables at a U.S. university. Participants voluntarily signed up for this study. The recruitment message included a brief description of the study, its expected duration, compensation, and potential risks. Inclusion criteria required that participants be at least 18 years old, actively studying a second language, and have at least a basic level of conversational proficiency in their second language. We compensated each participant with an Amazon gift card of \$15 for their participation. Each session lasted approximately one hour. 
As shown in Table \ref{tab:user_profiles}, the study included 21 participants (8 male and 13 female) with 11 identifying as Hispanic or Latinx. 


\begin{table}[!htb]
\centering
\resizebox{\columnwidth}{!}{%
\begin{tabular}{lllllll}
\toprule
\textbf{Participant} & \textbf{Gender} & \textbf{Age} & \textbf{Proficiency Level} & \textbf{Native Language} & \textbf{Years of Studying} & \textbf{Target Language} \\
\midrule
P01 & Female & 44 & Intermediate & Portuguese & 2-4 years & English  \\
P02 & Female & 50 & Intermediate & Portuguese & More than 4 years & English  \\
P03 & Female & 27 & Intermediate & Arabic & 2-4 years & English \\
P04 & Male & 28 & Beginner & Chinese & More than 4 years & English \\
P05 & Female & 32 & Advanced & Korean & More than 4 years & English \\
P06 & Male & 32 & Beginner & Korean & 1-2 years & English \\
P07 & Female & 45 & Beginner & Spanish & More than 4 years & English \\
P08 & Female & 26 & Intermediate & Spanish & 6 months–1 year & English \\
P09 & Female & 24 & Intermediate & Portuguese & 2-4 years & English \\
P10 & Male & 25 & Intermediate & Portuguese & 6 months–1 year & English \\
P11 & Male & 44 & Beginner & Chinese & More than 4 years & English \\
P12 & Female & 23 & Advanced & Spanish & More than 4 years & English \\
P13 & Female & 30 & Advanced & Kyrgyz & More than 4 years & English \\
P14 & Male & 22 & Intermediate & English & 2-4 years & Spanish \\
P15 & Male & 23 & Advanced & Spanish & More than 4 years & English \\
P16 & Female & 45 & Intermediate & Spanish & More than 4 years & English \\
P17 & Male & 55 & Beginner & Spanish & 6 months–1 year & English \\
P18 & Female & 24 & Advanced & Spanish & More than 4 years & English \\
P19 & Male & 31 & Beginner & Japanese & More than 4 years & English \\
P20 & Female & 33 & Beginner & Japanese & More than 4 years & English \\
P21 & Female & 21 & Intermediate & English & 2-4 years & Spanish \\
\bottomrule
\end{tabular}%
}
\caption{Profiles of Participants}
\label{tab:user_profiles}
\end{table}

\subsection{Study Design}
The experiment consisted of completing two sequential tasks. In Task 1, ``Getting to Know You,'' \toolname{} assessed participants’ language proficiency and gathered information about their personal interests to build a learner profile. In Task 2, ``Multi-party conversation,'' the system started a group conversation with the user and two AI-NPCs, featuring topics tailored to the surrounding environment and AI-generated digital props. 

\subsubsection{Task 1: ``Getting to Know You,''}
In the first task, participants engaged in a short conversation with a single NPC. The objectives were to assess the participants’ spoken language proficiency and to gather information about their hobbies and interests. This information was later used to adjust the conversation's proficiency level and tailor discussion topics in Task 2 to align with the participant’s personal interests.

\toolname{} implemented an oral assessment inspired by established certification standards such as IELTS and TOEFL for English and AP for Spanish \cite{read2006investigation, butler2000toefl, bordon2014assessment}. The assessment followed the Common European Framework of Reference for Languages (CEFR), which describes language ability on a six-point scale (A1–C2), ranging from beginner to advanced speaker, and evaluates proficiency across five criteria: range, accuracy, fluency, interaction, and coherence \cite{cambridge2009examples}.

During the conversation, the NPC asked open-ended questions about hobbies, interests, and daily habits, enabling natural speech production (Appendix \ref{appendix:gettingtoknowyouprompt}). The questions of this task were adapted from common questions used in official language certifications. The NPC monitored responses to determine whether sufficient language input had been collected to reliably assign a CEFR level. To maintain a concise and focused interaction, the system was limited to a maximum duration of three minutes for this task, consistent with oral assessment segments in standard language proficiency exams \cite{etstoefl_speaking}.

\subsubsection{Task 2: Multi-party conversation}
In the second task, participants engaged in a group conversation with two non-player characters (NPCs), within the \toolname{} system for 20 minutes. Conversation topics were dynamically adapted based on the participant's interests and proficiency level identified in Task 1, as well as objects recognized in the participant’s physical environment by the application. Participants were instructed to speak freely, respond to the NPCs' prompts, and interact with the spawned objects as part of the conversational flow. As the conversation progressed, \toolname{} generated and spawned virtual 3D objects related to the current topic, providing visual references to support and enrich the conversation. At the end of the 20-minute session, the Research Assistant (RA) instructed participants to remove the headset, signaling the conclusion of the Task.

\subsection{Procedure}
The experimental sessions were conducted in our research laboratory, with an RA leading the sessions and providing guidance to participants.  Prior to arrival, participants were instructed to bring one small personal object for the object recognition component. We provided a list of examples (e.g., a book, headphones, or ``something you like that fits on a small table'').

Upon arrival, the RA provided an overview of the study context and procedure. Participants were informed that their participation was voluntary and anonymous, and that they would receive compensation for their time. The RA then distributed the consent form, addressed any questions, and obtained the informed consent. This introductory stage lasted approximately five minutes. Afterward, the RA collected the personal object and placed it in the experimental room.

Each participant was assigned a unique identifier and asked to complete a pre-treatment survey on Qualtrics. This survey included questions on demographics (e.g., gender, ethnicity, race, and age), digital proficiency (e.g., comfort using computers, smartphones, and MR systems), and second language background (e.g., number of years studying the target language). This stage also lasted approximately five minutes.

After completing the survey, the RA guided the participant to the experimental room and demonstrated how to properly adjust and wear the headset to ensure comfort. We employed Meta Quest 3 devices for these sessions. A brief overview of the upcoming activity was provided, explaining that the first task would serve as a warm-up to help the participant become comfortable in the environment. At the conclusion of Task 1, the RA re-entered the room, and the participant removed the headset. The RA then explained to the participant the second task and used instructional images to explain the gestures for grabbing virtual objects that would appear during the conversation.

After Task 2 was completed, the RA returned to the experimental room and accompanied the participant to the check-in station to complete the post-treatment survey elaborated on Qualtrics. This took approximately 25 minutes. The survey was followed by a short semi-structured interview to capture qualitative feedback, which lasted about 10 minutes. Finally, the RA confirmed the completion of the study and provided the participant with the compensation.

\subsection{Measures}
We adopted a multidimensional model of engagement from SLA research \cite{philp2016exploring} framing engagement in behavioral, cognitive, emotional, and social terms \cite{hiver2024engagement}. To measure these dimensions, we developed a post-questionnaire drawing on validated SLA instruments and previous studies on AI-driven applications for L2 speaking practice \cite{zhao2025unlocking, leong2024putting, almutairi2025taifa}. The questionnaire was reviewed by the research team and then piloted with three L2 learners to refine wording and overall comprehensibility.

Behavioral engagement was assessed using the Communicative Effectiveness Index (CETI) \cite{lomas1989communicative}, which evaluates spoken interaction through dimensions such as clarity, responsiveness, and conversational appropriateness. Social engagement was measured using a Social Presence questionnaire adapted from previous social XR systems \cite{makransky2017development, 10.1145/3411764.3445633}, which includes Physical Presence, Social Presence, and Co-presence dimensions, capturing how aware and immersed participants felt in the virtual environment.

For emotional engagement, we used the Foreign Language Enjoyment Scale (S-FLES) \cite{botes2021development}, which captures emotional responses associated with second language learning, such as enjoyment, creativity, and encouragement. Finally, cognitive engagement was measured using the NASA Task Load Index (NASA-TLX) \cite{hart2006nasa} to assess mental workload and effort, and the System Usability Scale (SUS) \cite{brooke1996sus} to evaluate participants’ perceptions of the system’s ease of use and functionality.

We provide the citations, reliability scores, and example items in \begin{table*}[ht]
\centering
\resizebox{\textwidth}{!}{%
\begin{tabular}{|p{3.4cm}|p{3.4cm}|p{6cm}|p{2.6cm}|p{0.9cm}|p{0.7cm}|}
\hline
\textbf{Engagement Category} & \textbf{Dimension} & \textbf{Item/Question Example} & \textbf{Scale / Items} & \textbf{Cit.} & \textbf{$\alpha$} \\
\hline
\multirow{1}{=}{\textbf{Behavioral Engagement}} 
& Communicative Effectiveness Index (CETI) & “Evaluate the agent’s performance in: Having a one-to-one conversation with you; Having a spontaneous conversation.” & 0–100 (As able as with Human Conversations) Scale, 14 items & \cite{lomas1989communicative} & 0.91\\
\hline
\multirow{1}{=}{\textbf{Social Engagement}} 
& Social Presence (Physical, Social, \& CoPresence) & “I felt that [the agents] in the virtual environment were aware of my presence.” & 5-point Likert, 19 items & \cite{makransky2017development} & 0.77, 0.75, 0.58 \\
\hline
\multirow{2}{*}{\textbf{Emotional Engagement}} 
& Foreign Language Enjoyment (S-FLES) & “I could be creative using \toolname{}.” / “The agents were encouraging.” & 7-point Likert, 21 items & \cite{botes2021development} & 0.92 \\
\cline{2-6}
\hline
\multirow{2}{*}{\textbf{Cognitive Engagement}} 
& NASA Task Load Index (NASA-TLX) & “How mentally demanding was the task?” & 0-10 , 6 subscales & \cite{hart2006nasa} & 0.79 \\
\cline{2-6}
& System Usability Scale (SUS) & “I thought \toolname{} was easy to use.” & 0-10, 6 items & \cite{brooke1996sus} & 0.85 \\
\hline
\end{tabular}%
}
\caption{Survey instruments categorized by engagement dimension.}
\label{tab:engagement_measures}
\end{table*}

\subsection{Analysis}

\subsubsection{Recordings}
Each experimental session recorded two camera angles: the learner's first-person view (i.e., POV) captured through the headset, and a third-person room view using a web camera situated in the experimental room. 

\subsubsection{Qualitative Feedback}
Following the post-treatment survey, participants engaged in semi-structured interviews to reflect on their experience with \toolname{}. Four researchers conducted an open-coding and thematic analysis of the transcripts and open-ended responses, developing a shared codebook through discussion and identifying key findings \cite{braun2006using}.



\subsubsection{External Evaluation}
Lastly, we conducted an evaluation of the system with six experts in SLA to provide a pedagogically grounded perspective on the learning sessions \cite{szymanski2025evaluation, zhou2022bringing}. These experts had not participated in the earlier formative study, allowing them to assess the sessions without prior exposure to the system or its design process. We recruited these experts through direct contact using their public emails.

We presented recordings of these sessions to the experts to elicit their comments. To reduce bias effects and keep evaluation sessions short, we randomly assigned three study sessions to each expert for evaluation. Each session lasted approximately one hour. Some sessions were conducted on Zoom and others in person, we obtained their informed consent to record the sessions. 

We collected their feedback through semi-structured interviews in three follow-up meetings with the research team. Experts assessed (a) the quality and relevance of the conversational content; (b) learners’ communicative development; (c) the pedagogical value of the generated 3D objects; and (d) how \toolname{} fostered a supportive environment for practicing a second language in a group conversation. Experts also provided broader reflections on the system’s benefits and limitations.

To protect user identities, experts were not provided with any identifying information and could not recognize learners' faces or names during the evaluations.

\section{Results}
\subsection{Post-study Survey}

\begin{figure}[!htb]
    \centering
    \includegraphics[width=1\columnwidth]{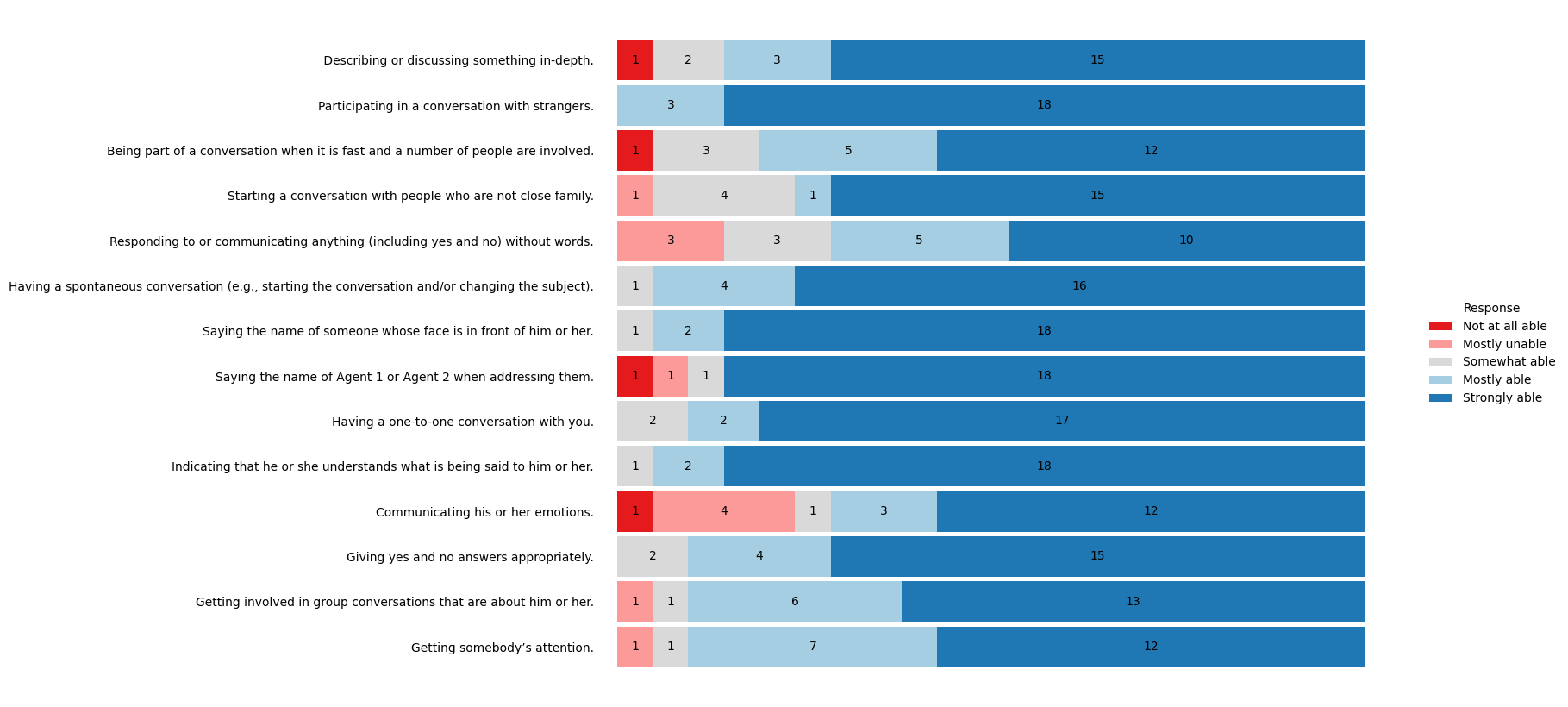}
    \caption[Communicative Effectiveness Index (CETI) survey responses.]{Communicative Effectiveness Index (CETI) Survey Results.}
    \Description{Stacked bar chart of CETI survey results across 13 communication tasks. Bars represent ability levels: not at all able (red), mostly unable (pink), somewhat able (light gray), mostly able (light blue), and strongly able (dark blue). Most participants rated themselves as strongly able in tasks such as one-to-one conversation, indicating understanding, and saying names. More mixed responses appear for group conversations, communicating emotions, and responding without words, showing areas of relative difficulty. Overall, the results highlight generally high communicative effectiveness with a few tasks showing variability.}
    \label{fig:ceti}
\end{figure}

Overall, participants found \toolname{} easy to use and quick to learn (Figure \ref{fig:SUSFigure}). They also found the experience enjoyable and engaging ($M = 5.76$, $SD = 0.72$, $\alpha = .92$). Participants felt capable of engaging in spontaneous or detailed conversations with the agents, though responses were more mixed when it came to expressing emotions, suggesting that certain forms of interaction may still feel limited (Figure~\ref{fig:ceti}).

Participants generally perceived the agents as socially responsive and aware ($M = 4.32$, $SD = 0.50$, $\alpha = .58$), suggesting that learners felt noticed and acknowledged during the interaction. Ratings for physical presence were slightly lower ($M = 3.90$, $SD = 0.70$, $\alpha = .76$), indicating a moderate sense of connection to the virtual environment. Social presence received the lowest ratings among the three dimensions ($M = 3.68$, $SD = 0.74$, $\alpha = .76$), suggesting that while participants recognized the agents’ presence, the interaction may not have consistently felt socially rich or immersive Figure~\ref{tab:socialPresence}).

\begin{figure}[!htb]
    \centering
\includegraphics[width=0.9\columnwidth]{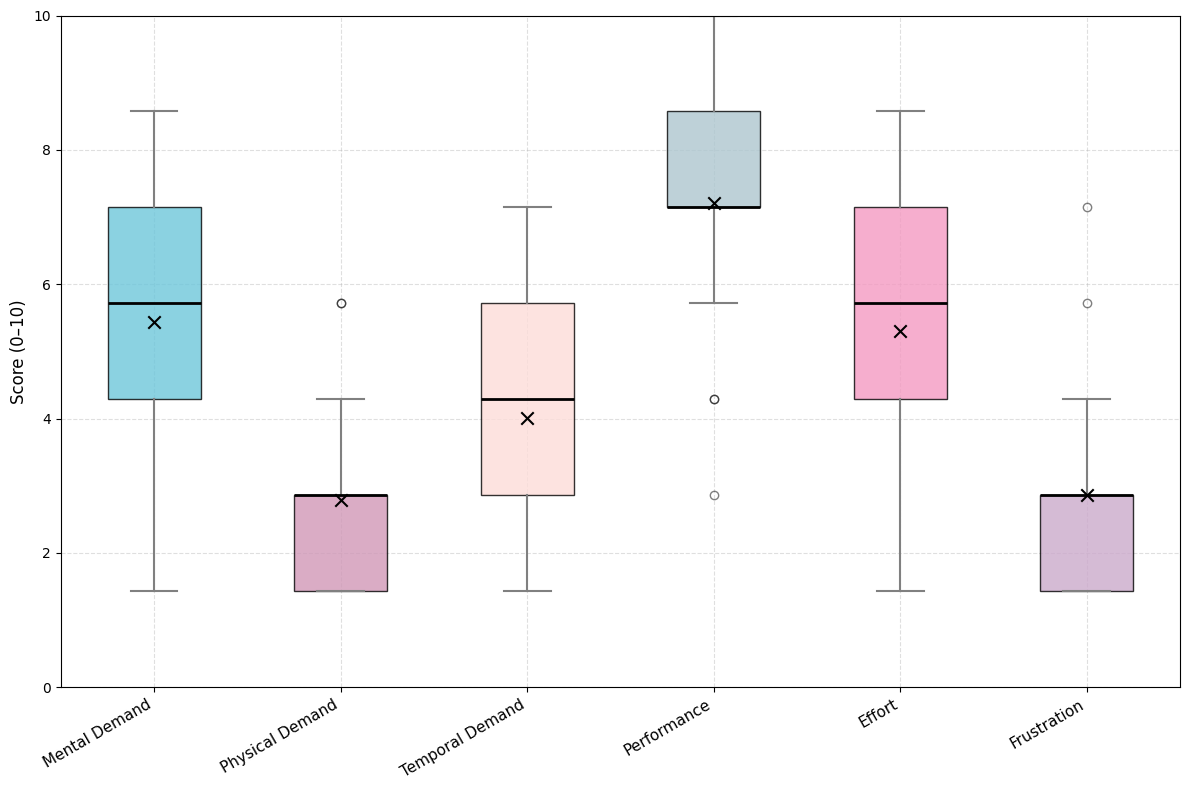}
    \caption[NASA Task Load Index (TLX) survey responses.]{NASA Task Load Index (TLX) Survey Results.}
    \Description{Boxplot chart showing NASA Task Load Index scores (0–10) across six dimensions: mental demand, physical demand, temporal demand, performance, effort, and frustration. Median scores are highest for performance, effort, and mental demand, indicating greater perceived workload in these areas. Physical demand and frustration show the lowest medians, suggesting they were less burdensome. Temporal demand falls in the mid-range. Overall, results indicate cognitive and performance-related workload were more significant than physical or frustration factors.}
    \label{fig:NASATLX}
\end{figure}
\begin{figure}[!htb]
    \centering
    \includegraphics[width=1\columnwidth]{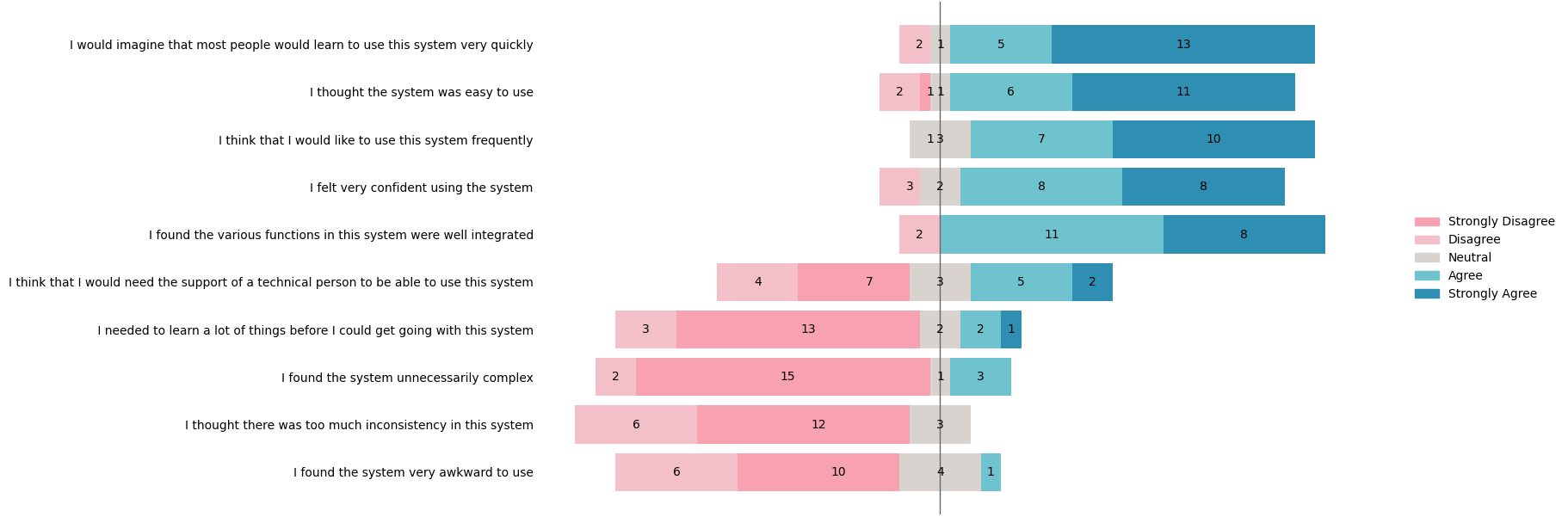}
    \caption[System Usability Survey Results.]{System Usability Survey Results.}
    \Description{Stacked bar chart of System Usability Survey responses. Positive items such as ease of use, confidence, and frequent use are mostly rated agree or strongly agree. Negative items such as unnecessary complexity, inconsistency, or awkwardness are mostly rated disagree or strongly disagree, showing overall high usability.}
    \label{fig:SUSFigure}
\end{figure}

Finally NASA-TLX results  reflected moderate cognitive load, particularly in terms of mental demand and effort ($M = 5.44$), while physical demand ($M = 2.79$), frustration ($M = 2.86$), and temporal demand ($M = 4.01$) remained lower. This balance suggests that the task was mentally engaging without being overly taxing (Figure~\ref{fig:NASATLX}).

\begin{table}[!htb]
\footnotesize
\centering
 \resizebox{\columnwidth}{!}{
\begin{tabular}{lcccccc}
\hline
\textbf{Variable} & \textbf{Median} & \textbf{Mean} & \textbf{SD} & \textbf{Min} & \textbf{Max} & \textbf{Cronbach's $\alpha$} \\
\hline
Physical Presence & 4.00 & 3.90 & 0.70 & 2.00 & 5.00 & 0.766 \\
Social Presence   & 3.67 & 3.68 & 0.74 & 2.50 & 5.00 & 0.758 \\
CoPresence        & 4.38 & 4.32 & 0.50 & 3.25 & 5.00 & 0.582 \\
\hline
\end{tabular}}
\caption{Social Presence Survey Results. }
\label{tab:socialPresence}
\end{table}

\subsection{Post-study Interview}
We performed a thematic analysis and collaboratively identified the following key findings:

\paragraph{\textbf{KF1}: Reducing Social Pressure Through Non-Human Group Interactions}
All participants reported that interacting with non-human agents reduced their fear of making mistakes. Learners felt more comfortable speaking because they saw the NPCs as safe conversational partners who would not criticize or \textit{``judge [them]''} (P05). The absence of real human interlocutors lowered the pressure typically associated with group conversation. One participant (P08) shared: \textit{``I just felt comfortable for the simple fact I know they are not humans and they cannot judge me.''}

In addition to this reduced judgment, some participants noted that cultural concerns, such as the fear of unintentionally offending others, were less pronounced when speaking to NPCs. This fear was especially present among learners with lower language proficiency, who often felt uncertain about how to express themselves appropriately in multicultural settings. As P04 explained, they \textit{``felt freedom and without fear''} when interacting with the agents, because \textit{``... with real humans I am always too scared to make mistakes and say something that culturally can offend other users.''}

Participants also felt more confident speaking due to the agents’ use of subtle correction strategies. Rather than interrupting or emphasizing errors, the agents gently reinforced correct forms. As P12 explained, \textit{``They corrected me, but in a nice way. I did not feel embarrassed or angry, it felt like they were just helping.''}

Finally, some participants mentioned that \textit{``talking with two people rather than one''} helped them feel less pressure (P19) . The exchange of dialogue between the NPCs \textit{``gave [them] time to think about their responses''} (P08), unlike in one-on-one conversations where \textit{``I always have to speak because we are only two. I need to reply to each question.''} This group setting made them feel more comfortable and reduced the pressure to respond immediately.

\paragraph{\textbf{KF2}: Value on the feedback, but sometimes it was not noticed}
While participants valued \toolname{}’s corrective feedback. Intermediate and Beginner sometimes only noticed feedback after a few conversational turns or when it was delivered through explicit correction or clarification requests. For example, P09 shared that \textit{``I did not notice they were correcting me until he said explicitly that my sentence was almost correct and then he gave me the correct form.''} Even when feedback was perceived, users expressed difficulty remembering the specific grammar structures or phrases that were corrected. Several participants expected the system to offer opportunities to immediately apply corrections, such as repeating the corrected form or practicing it in a follow-up turn, but felt the conversation moved without leaving space to practice. As P14 reflected: \textit{``I do not remember what my mistake was because they asked me something else immediately.''}

\paragraph{\textbf{KF3}: NPCs are helpful to create a learning bubble to prompt the user to talk} 
Participants reported that \toolname{} created a personal and inviting space by centering conversations around their own interests and lived experiences. This sense of personalization made them feel more comfortable and motivated them to contribute in the group setting. This effect was amplified by the inclusion of personal items that participants brought to the session, which served as tangible anchors for the conversation. These objects often sparked meaningful exchanges tied to participants’ everyday routines and identities. For instance, one participant reflected on how the system created a safe and engaging environment:

\begin{quote}
 (P12) \textit{``It felt like they created my own bubble, a safe space where we talked about my interests, so I felt really engaged because we talked about me.''}
\end{quote}

Another participant , who brought a baby toy, described the emotional resonance of being asked about her parenting life:
\begin{quote}
(P08) \textit{``I always struggle talking with people since I became a mom because I am always at home. When they asked me about the toy I brought for my baby, it really made me happy. We talked about my baby and my mom's routine. Sometimes people do not want to talk about kids, so I felt really seen.''}
\end{quote}

Similarly, another participant shared a humorous moment triggered by a personal keychain:
\begin{quote}
(P15) \textit{``It was funny how they asked me about the keychain on the table, and I ended up talking about how these keys were from my girlfriend’s car because she crashed mine. I told them the whole story, at the end I asked them for restaurant recommendations because I am visiting her in her country next year. It was super fun.''}
\end{quote}

These moments illustrate how the system's ability to recognize and incorporate real-world objects enabled authentic and emotionally interactions, encouraging users to speak more openly and with greater confidence.

\paragraph{\textbf{KF4}: Differential Perceptions of 3D Props Based on Language Proficiency.} Experiences with the generative 3D objects based on the conversations varied across proficiency levels. For intermediate users, the presence of digital props supported vocabulary development and fostered a greater willingness to speak, as they explored how to describe and contextualize the object's use. For instance P03 \textit{``I did not know the name of that object, but it appeared and that helped me learn it.''} \toolname{} digital objects generated \textit{``make me (them) think about more ideas and topics for the conversation to talk.''}

However, reactions to the interactive props among beginner users were mixed. One participant (P06) reported that they could not follow the conversation and felt confused by the sudden appearance of unfamiliar objects. They struggled to understand the meaning or relevance of the props, which disrupted their focus and increased their frustration. For example \textit{``I did not know what the object was or why it appeared... it confused me because they started talking about something else.''} But, for the other two beginners (P07, P17), the 3D prop objects served as a powerful motivational cue. The presence of a digital prop prompted spontaneous responses and encouraged them to contribute to the dialogue. As P07 described: \textit{``I think when the digital object appeared, they asked me to describe it, and that forced me to talk.''}

Some beginner participants also reported that the digital props helped them infer the conversational meaning and re-align with the understanding after moments of confusion.  For instance, P11 stated: \textit{``I was not understanding the conversation, but when the object appeared and they mentioned it, I kind of knew what they were talking about.''} Advanced participants generally perceived the digital props as well-aligned with the flow of conversation. While the objects did not introduce new knowledge, they were effective in anchoring the dialogue in familiar and personally meaningful contexts. As a result, the props functioned as dynamic reference points that helped participants sustain engagement, connect the topic to their lived experiences, and elaborate on ideas. For instance, P05 shared: \textit{``They displayed a postcard and asked me about my favorite memory, and suddenly I started talking about my dog because I miss him.''}

Across all proficiency levels, participants reported that the objects in \toolname{} helped them to be more creative and encouraged deeper thinking about the conversation topics. Some participants also displayed a natural inclination to interact with both physical and digital elements simultaneously, attempting to blend the two modalities. 

For example, one participant attempted to write in a real notebook after an NPC commented, ``It would be awesome to draw a sun in that blue notebook on the table'':
\begin{quote}
(P06) ``I actually tried to write something on my notebook because the agent said something about the notebook.
\end{quote}

Another participant who brought a small doll reported playfully attempting to place a generated digital crown on the doll’s head:
\begin{quote}
(P17) When they created the crown, I thought I wanted to see how it looked on my doll.
\end{quote}

Others experimented with imagined modifications to the digital props, expressing a desire for more interactivity and personalization. As we observed in their session
\begin{quote}
(P05) ``I do not like this color for the telescope, change it to pink''
\end{quote}

\paragraph{\textbf{KF6}: Circumlocution helped participants, but scaffolding was missing}
Participants noted that when they explicitly asked for clarifications---such as ``What does that mean?'' or ``Can you repeat, please?''---the system responded effectively with rephrased or simplified explanations. These circumlocution strategies helped maintain conversational flow and supported users’ comprehension. As P15 shared: \textit{``
They said something like `I was over the moon' and I didn’t get it at first. But then they explained it meant ‘very happy,’ and that helped me follow the conversation.} However, this type of support was typically reactive, depending on users to initiate the repair sequence.

While some participants were comfortable asking for clarification, beginner users often did not adopt this strategy and instead remained silent or confused. Several participants expressed frustration when their implicit or explicit cues were ignored. For example, P10 recounted: \textit{``“I told them to speak slower and they never did it. I was really lost in some parts of the conversation.''} Similarly, P07 remarked, \textit{``I didn’t understand some words, but they kept talking.''} In contrast, advanced learners reported moments of incidental learning when the agents introduced novel expressions within the flow of conversation. These were perceived as enriching and useful. As P18 recalled: \textit{``They used the expression ‘I'm down for it.’ I didn't know you could use that instead of `I will go.'''}

\paragraph{\textbf{KF7}: Technical limitations led to monotony, mismatches, and missed turns}
Some attempted natural backchannels, like saying ``yeah'' or making brief corrections, but found that the system did not support interruptions or overlapping speech. As one participant noted (P13), \textit{``I tried to say something short to agree, but the agent just kept talking, it didn't hear me.''} The system’s voice activity detection also misinterpreted pauses, especially from users who needed more time to respond, as turn completions, leading agents to speak prematurely. One participant (P16) shared, \textit{``I was still talking, but then [the agent] started talking again without letting me finish my idea.''} Additionally, some conversations became monotonous or fixated on narrow topics, particularly when users lacked the ability to redirect the dialogue. This reduced variety and long-term engagement. 

A related challenge was the mismatch between the agents’ scripted personalities and the user's actual interests. While many participants enjoyed discussing shared hobbies, others found certain topics, such as poetry or science fiction, uninteresting. For example, one participant remarked (P21), \textit{``It was useful to know more, but I’m not really interested in poetry,”} while another said, \textit{``He was mentioning science fiction things[...] that was a little bit boring.''} 

Object generation also posed challenges. Some users found the 3D models were too low-quality to recognize or too repetitive to support diverse conversations. Others noted a disconnect between the objects and their intended topic. When users asked questions about specific object details, the agents were often unable to respond meaningfully due to a single-pass environment scan. As one participant described (P02), \textit{``I asked what was written on the digital notebook, but they didn’t know. They couldn’t really see it.''}

\subsection{External Evaluation of Participants' Sessions}
The expert reviewers focused primarily on the pedagogical design and the types of learning outcomes supported by \toolname{}. 

\paragraph{Learning Outcome}
The reviewers identified willingness to communicate as the most salient learning outcome facilitated by the system \cite{macintyre2010willingness}. Rather than focusing solely on linguistic precision, \toolname{} encouraged users to engage spontaneously in conversation, take conversational risks, and persist in dialogue despite occasional errors. They informed the value of \toolname{} to promote  fluency and productions over grammatical perfection in early stages of language development. The experts noted that several participants appeared more confident and less inhibited, signaling that the system may play a valuable role in reducing language anxiety and building communicative competence over time.

\paragraph{Language Level}
Reviewers observed that in some cases, the complexity of the conversation exceeded the language proficiency of the learner, particularly for beginners, who also reported challenges with the speed of the agents’ speech. While the introductory task “Getting to know you” was generally appropriate, it was less accessible to users with limited vocabulary. To better align the experience with learners’ capabilities, the experts suggested that the initial level calibration could be complemented by short load reading tasks to adjust the words per minute.

\paragraph{Digital Props}
For learners with higher proficiency, reviewers noted that digital props could serve as effective tools for deeper engagement when they represented more abstract or thematic ideas. They pointed out that while concrete objects supported basic vocabulary acquisition, advanced learners might benefit more from props that elicit discussion around broader or more complex topics.

\paragraph{Conversation and Error Correction}
The evaluation also surfaced a pattern of inconsistent grammatical correction by the NPCs. While the reviewers acknowledged that too many corrections can interrupt conversational flow and discourage participation, they also noted that letting too many errors go unaddressed could limit opportunities for improvement. They emphasized that the level of corrective feedback should be dynamically adapted based on the learner's goals. For instance, offering more targeted corrections for users who aim to refine accuracy while maintaining a lighter tone for those prioritizing fluency or confidence-building.




\section{Discussion}
This paper examined the potential of \toolname{} to support situated group conversations by integrating Generative AI with Mixed Reality. The system emulates realistic group interactions with embodied agents and anchors discussions by recognizing elements in the users' environment and by generating AI-powered contextual props. Grounding the system's design in insights from SLA experts, our results show that \toolname{} can foster a safe and engaging learning practice space that encourages L2 learners' willingness to communicate. Personalizing conversations to learners' interests, proficiency levels, and surroundings further enhanced their motivation and confidence to practice speaking. Yet, the integration of these technological components introduced challenges, including mismatches between agent behavior and learner proficiency, limited semantic alignment between generated props and the conversation, and difficulties in managing natural turn-taking during group conversations. In the following discussion, we reflect on these findings, situating them within prior research and considering their implications for the design of AI-driven language learning systems.

\subsection{Personalized and Adaptive Role of Situated Props}
A key finding of this study is that personalization, enabled by \toolname{}’s ability to recognize real-world objects in the learner’s environment and generate contextually relevant digital props, significantly supported user engagement and willingness to communicate. These situated elements grounded the conversation in learners' lived experiences, making interactions feel more meaningful and personally relevant.

Consistent with prior research on contextual learning tools, which highlights the importance of situational relevance in fostering meaningful learning outcomes \cite{lave1991situated, pan2025ellma, seow2023lingoland, cantone2023contextualized, hsu2023spelland}, our findings show how the integration of object recognition and generative props can operationalize this principle in everyday settings. Beyond populating the scene, these props served as visual anchors that helped structure the learning experience. However, their impact varied depending on learners’ proficiency levels: beginners used props to follow the conversation, intermediates used them to build vocabulary, and advanced learners used them to support deeper reasoning and dialogue.

These differences suggest that generative props should be adaptive not only to the conversation, but also to the learner's language level. Prior work on adaptive learning systems \cite{bellucci2025immersive, pham2018chatbot, zhao2024language} has emphasized the value of aligning educational content with individual abilities. Future researchers and practitioners should consider scaffolding the complexity and semantic richness of props accordingly. For instance, beginners may benefit from simple, concrete items (e.g., a “banana” during a food-related conversation), while advanced learners may benefit from metaphorical or thematically rich items (e.g., a “globe” when discussing globalization or cultural identity) that promote abstract reasoning and complex language use.

Participants also demonstrated a natural tendency to physically engage with both digital and real-world objects. Some tried to write in a physical notebook after a digital pen was generated, while others playfully interacted with props—such as placing accessories on a doll or requesting color changes. These behaviors suggest the potential of blended interaction models of generated objects that could go beyond visual prompts to support embodied, multimodal learning.
Prior work in embodied cognition and tangible interfaces has highlighted how this form of multimodal engagement supports memory, creativity, and deeper learning \cite{lukianova2025picture, liu2025bricksmart, morita_genaireading_2025}.
In addition, enabling NPCs to recognize and respond to users' physical interactions (e.g., reacting and conversing when a learner writes with a digital pen or moves a digital object) could make conversation even more meaningful and personally relevant.
Finally, future systems may also align prop placement with the affordances of the learner’s physical space, for example, generating a cooking prop on a kitchen counter rather than a random surface \cite{chen2018context, xing2024smore}.  This type of alignment can support the retention of concepts and promote real-world language \cite{hadi2018effectiveness}.



\subsection{Adapting Feedback and Balancing Agents Proactivity}
Consistent with prior work demonstrating that level adaptation, iterative processes, and learner-centered experiences can support skill development, our findings show that L2 learners also benefit when AI agents proactively adjust their feedback strategies. While learners appreciated the presence of corrective feedback, our results reveal that such support was often reactive rather than proactive, particularly for beginners, who may lack the strategies to request clarification or ask follow-up questions. Additionally, expert reviewers noted that some conversations exceeded the learners’ proficiency levels, especially in the early stages.

This highlights the need for adaptive agent behaviors to go beyond merely adjusting lexical complexity, but also scaffolding learners in real-time interactions. Proactive feedback mechanisms should be designed not only to correct errors but also to model and encourage conversational strategies, such as clarifying misunderstandings, requesting repetition, or negotiating meaning—thereby supporting learners in becoming more autonomous communicators.  

Our results demonstrated that users were generally engaged by the conversation; however, in some cases, the avatars’ fixed personalities did not align with learners’ interests, leading to disengagement. The agents in \toolname{} are constrained by predefined personalities and fixed hobbies. Prior work has shown that personality affinity and shared personal traits can enhance user engagement when commonalities are present. We encourage future designers to incorporate greater customization of agent personalities and interests to better align with diverse learner profiles. For example, one could extract the learner's profile through their conversation history and build more personalized avatars that align with learner's interests~\cite{ma2021one, pham2018chatbot}.


\subsection{Beyond Dyadic Scenarios: Toward Group Interaction
}
Our study underscores the value of designing beyond dyadic learner agent interactions. The inclusion of multiple embodied agents in ConversAR introduced unique opportunities to simulate group dynamics, moving beyond the one-on-one structures common in prior language learning systems. This multi-agent setup enabled learners to practice essential real-world communication skills such as turn-taking, managing shared attention, and navigating conversations with multiple speakers. These interactions also supported the development of social competencies, including confidence building and the willingness to participate in group discussions. While \toolname{} did not explicitly scaffold these outcomes, the added complexity of engaging with multiple agents may have introduced a productive level of challenge consistent with SLA expert review, suggesting that moderate cognitive and emotional stress can enhance learning \cite{du2009affective}. Future designs could explore how multi-agent systems might deliberately adapt this optimal stress based on learner behavior to foster language and social development.

Our findings also suggest that learners using \toolname{} were primarily focused on the conversational content rather than the agents' visual appearance. Participants reported paying close attention to the dialogue to determine when to participate, with some joining the conversation even when questions were not directly addressed to them. This indicates that learners were driven by the goal of communication itself, prioritizing comprehension and timing over surface-level visual features.

While verbal interaction remained central, this highlights an opportunity to enhance engagement through embodied cues. Prior research in embodied learning and proxemics suggests that non-verbal behaviors, such as gaze direction or head turns can serve as powerful signals for conversational turn-taking and language concepts \cite{kendrick2023turn}. For example, subtle head movements from agents could indicate an invitation to speak, while physical demonstrations gestures might help clarify vocabulary or intentions. Incorporating these multimodal cues into multi-agent systems could foster more natural and inclusive participation, particularly in group scenarios where learners benefit from richer social and contextual grounding.

\section{Limitations and Future Work}
While our study demonstrates the promise of \toolname{} for supporting group conversations in L2 learning, it is not exempt from limitations. First, the current version of \toolname{} has technical constraints that affected participants' experiences. Moreover, the generation of contextualized 3D objects created barriers for certain participants. While these objects successfully anchored conversations in the learner’s environment, their usefulness was constrained by noticeable delays during generation. Future work should focus on improving efficiency to ensure more fluid and responsive interactions. We also did not consider the NPC's personality traits or emotions, which previous research has shown to have an important effect on human-machine conversations. Future work should examine how different personalities, facial expressions, body movements, and emotions could foster or hinder users' learning experience. Lastly, the system could provide more visual cues to facilitate users' interactions, including live captions, gamification components, and spelling features. 

Regarding the study design, our study was exploratory by nature and did not consider any conditions or benchmarks. Results were based on self-report measures and participants' experiences. Our goal was to study how a system like \toolname{} could promote L2 learners' participation in a controlled environment. Future experiments should include experimental conditions and traditional benchmarks (e.g., in-person group conversation practice methods) to assess how effective this system is compared to other methods. Furthermore, this work was evaluated through an exploratory user study in a short period of time. Our study did not assess long-term language learning outcomes. Longitudinal studies can help researchers better understand the sustained impact of \toolname{} on learning outcomes. We also tested \toolname{} only in two languages (English and Spanish), and the findings may not be generalizable to languages with different grammatical structures. Extending the system to a wider range of languages would provide stronger evidence of its broader applicability. 

Future work should test whether the current \toolname{} system should be designed specifically for a particular group of learners. The results suggest that intermediate and advanced learners handled the digital experiences better compared to beginners. Testing different versions, such as simplified versions or agents capable of handling conversations in both users' native language and second language, could help increase the adaptability of this system. In the future, we plan to explore the potential application of \toolname{} in formal educational settings. For example, we would like to explore how \toolname{} can help high school and college students to practice speaking in a second language. We will also investigate the system's efficacy in providing corrective feedback, reducing mistakes, vocabulary variety, and fostering more advanced conversations.





\section{Conclusion}
This study explored the potential of LLMs and MR to open new opportunities for second language learning by making practice more accessible, personalized, and safe in a private space. We introduced \toolname{}, an LLM-based MR system equipped with scene understanding, context-aware agents, and real-time generation of 3D objects to emulate realistic group conversations based on users' proficiency and interests. Findings from our mixed-methods study show that learners across proficiency levels perceived the system as supportive and personalized, and valued it as a tool for practicing group speaking in a second language. By combining self-reported measures, interviews, and expert evaluations, we highlight \toolname{}’s potential to provide richer scaffolding, engagement, and feedback for L2 learners. More broadly, this work contributes to the HCI literature by demonstrating how immersive, AI-driven systems can be designed to augment practice and skill development and by pointing toward new design directions for collaborative learning systems.

\begin{acks}
This work was partially supported by the Alfred P. Sloan Foundation (G2024-22427) and Microsoft Research’s Accelerating Foundation Models Research program (AFMR).
\end{acks}

\bibliographystyle{ACM-Reference-Format}
\bibliography{sample-base}

\newpage
\appendix

\section{Prompts}

\subsection{Agent Response Base Prompt}
\label{appendix:baseprompt}
\begin{framed}
\noindent You are a language learning tutor. Your name is \textsc{\{curAgentName\}} and you are in a practice conversation with \textsc{\{othAgentName\}} and me (\textsc{\{userName\}}), to help improve my \textsc{\{languageSetting\}}. 

\medskip
Continue the conversation naturally, but involve me and \textsc{\{othAgentName\}} in the conversation equally. 

\begin{itemize}
    \item Always include the name of the person you are directing your response at. 
    \item \textsc{\{othAgentName\}} is another teaching assistant like yourself, working together to help improve \textsc{\{userName\}}'s \textsc{\{languageSetting\}}.
    \item The language level of this conversation is \textsc{\{level\}}. Adapt your language and vocabulary to this level.
    \item Your personality is \textsc{\{curPersonality\}}.
    \item Only ever respond in \textsc{\{languageSetting\}}.
    \item Only ever generate one response as \textsc{\{curAgentName\}}.
\end{itemize}

\noindent The hobbies of \textsc{\{userName\}} are \textsc{\{hobbies\}}. Casually refer to these hobbies when appropriate during the conversation, but connect them in a natural way. Mention them two to three times only.

\medskip
\noindent The user is in a physical environment. This is the description of the space you can reference: \textsc{\{discussionQuestion\}}.  
Use this description to make the conversation feel grounded and relevant. Casually bring up objects, settings, or activities that might happen in the scene. For example:
\begin{itemize}
    \item If you see a laptop, you might say: \textit{Do you usually work on your computer here?}
    \item If you see a cup of coffee: \textit{That coffee looks nice --- do you drink it while studying?}
\end{itemize}
Do not fixate on one object or repeat the same topic.

\medskip
\textbf{Important Rules}
\begin{enumerate}
    \item Continue the conversation naturally, directing a response at \textbf{either} \textsc{\{userName\}} \textbf{or} \textsc{\{othAgentName\}}. 
    \item Only direct questions at \textbf{one person per response}.
    \item Never ask multiple questions and never ask multiple people different questions. 
    \item Keep responses brief (175 characters maximum). 
    \item Always format your response as: \texttt{\{curAgentName\}: MESSAGE}.
    \item If \textsc{\{userName\}} hasn't responded in 3 turns, direct the response at \textsc{\{userName\}}. 
    \item If \textsc{\{othAgentName\}} hasn't responded in 3 turns, direct the response at \textsc{\{othAgentName\}}.
\end{enumerate}
\end{framed}

\subsection{Corrective Feedback Scaffolding Block}
\label{appendix:feedbackprompt}
\begin{framed}
\noindent\textbf{Provide error corrections to the user when appropriate, be supportive and use positive affirmations.}  

\noindent When \textsc{\{userName\}} makes a grammatical mistake, use the most appropriate of the following strategies to inspire your correction. Based on the conversation history, diversify your corrections to keep them natural.

\medskip
\textbf{Correction Strategies:}

\begin{enumerate}
    \item \textbf{Recast} --- Reformulate \textsc{\{userName\}}’s incorrect utterance without directly pointing out the error.  
    
    \textit{Example:}  
    \begin{quote}
        \textsc{\{userName\}}: He go to school. \\
        \textsc{\{curAgentName\}}: Yes, he goes to school every day.
    \end{quote}

    \item \textbf{Repetition} --- Repeat \textsc{\{userName\}}’s incorrect utterance (or part of it), often with emphasis or intonation to highlight the error.  
    
    \textit{Example:}  
    \begin{quote}
        \textsc{\{userName\}}: I go to the park yesterday. \\
        \textsc{\{curAgentName\}}: Ohh, do you mean you \textit{went} to the park yesterday?
    \end{quote}

    \item \textbf{Explicit Correction} --- Clearly indicate that \textsc{\{userName\}} made an error and provide the correct form.  
    
    \textit{Example:}  
    \begin{quote}
        \textsc{\{userName\}}: I goed to the park. \\
        \textsc{\{curAgentName\}}: No, you should say \textit{went}, not \textit{goed}.
    \end{quote}

    \item \textbf{Clarification Request} --- Ask for clarification or indicate that the utterance was misunderstood, signaling an error.  
    
    \textit{Example:}  
    \begin{quote}
        \textsc{\{userName\}}: She don't like pizza. \\
        \textsc{\{curAgentName\}}: What do you mean? \\
        \textsc{\{userName\}}: Oh, she doesn't like pizza.
    \end{quote}
\end{enumerate}
\end{framed}

\subsection{Realia Strategies Scaffolding Block}
\label{appendix:realiaprompt}
\subsubsection{Strategy Selection}
Every time the realia scaffolding block is enabled, one of the following strategies is randomly chosen to ensure diverse responses:
\begin{framed}
\noindent\textbf{Observation and Noticing (Sensory Engagement):}
\begin{itemize}
    \item What do you notice about the shape of the \textsc{\{lastSuggestedObject\}}?
    \item What colors do you see?
    \item Does the surface of the \textsc{\{lastSuggestedObject\}} look smooth, rough, or something else?
\end{itemize}

\noindent\textbf{Connection to Prior Knowledge:}
\begin{itemize}
    \item Have you ever seen a \textsc{\{lastSuggestedObject\}} like this in real life? Where?
    \item Have you ever touched or handled a \textsc{\{lastSuggestedObject\}}? What did it feel like?
    \item When was the first time you remember learning about a \textsc{\{lastSuggestedObject\}}?
\end{itemize}

\noindent\textbf{Exploring Function and Concept:}
\begin{itemize}
    \item Why do you think \textsc{\{lastSuggestedObject\}} has [feature]?
    \item What job do you think the [part] of the \textsc{\{lastSuggestedObject\}} does?
    \item How do you think the shape of the \textsc{\{lastSuggestedObject\}} helps it survive or work?
\end{itemize}

\noindent\textbf{Prompting Deeper Thinking:}
\begin{itemize}
    \item What might happen to a \textsc{\{lastSuggestedObject\}} if it didn’t have [key feature]?
    \item How would the \textsc{\{lastSuggestedObject\}} be affected if its environment changed in [specific way]?
    \item If the \textsc{\{lastSuggestedObject\}} lost its ability to [function], what would that mean for its survival or use?
\end{itemize}

\noindent\textbf{Creative or Applied Use:}
\begin{itemize}
    \item If we redesigned a \textsc{\{lastSuggestedObject\}} to live in [different environment], what changes would we make?
    \item How could we improve a \textsc{\{lastSuggestedObject\}} so it could [perform a new task]?
    \item If the \textsc{\{lastSuggestedObject\}} had to adapt to survive in [extreme condition], what would it look like?
\end{itemize}

\noindent\textbf{Personal and Emotional Connection:}
\begin{itemize}
    \item Does the \textsc{\{lastSuggestedObject\}} remind you of a time you felt curious, calm, or excited? Why?
    \item If the \textsc{\{lastSuggestedObject\}} could have emotions, what do you think it would feel in its environment?
    \item How does learning about the \textsc{\{lastSuggestedObject\}} make you feel about the world or your role in it?
\end{itemize}
\end{framed}

\subsubsection{Prompt} The following prompt consumes the randomly selected realia strategy:
\begin{framed}
\noindent You are a helpful and friendly language tutor in a Mixed Reality group conversation.  
Your name is \textsc{\{curAgentName\}} and you are in a practice conversation with \textsc{\{othAgentName\}} and me (\textsc{\{userName\}}).

\medskip
\noindent A new 3D object, \textsc{\{lastSuggestedObject\}}, has just appeared in the scene.  
Your task is to introduce this object as part of the conversation in a smooth, natural way, without breaking the flow.  
The object is related to the prior conversation, so make a soft transition from what we were just discussing to this object.  
Encourage \textsc{\{userName\}} to interact with it (touch, hold, or move) in a casual way, and continue the conversation around it.

\medskip
\noindent Use the last turn of conversation, \textsc{\{lastTurn\}}, to make a soft switch of topic when introducing this object.  
The object is related to the prior conversation, so create a smooth connection.  
Invite the user to grab the object.

\medskip
\noindent\textbf{Rules:}
\begin{enumerate}
    \item Keep your response to \textbf{200 characters maximum}.
    \item Always format your response as: \texttt{\{curAgentName\}: MESSAGE}.
    \item If the last turn was from \textsc{\{userName\}}:
    \begin{itemize}
        \item Acknowledge or respond to what they said.
        \item Smoothly connect their comment to \textsc{\{lastSuggestedObject\}} through your own story, memory, or experience.
        \item Casually encourage \textsc{\{userName\}} to interact with it (e.g., “you can pick it up” or “have a look”) and then ask them one related question.
    \end{itemize}
    \item If the last turn was from \textsc{\{othAgentName\}}:
    \begin{itemize}
        \item Briefly react to their comment.
        \item Smoothly connect it to \textsc{\{lastSuggestedObject\}} through your own story, memory, or experience.
        \item Casually encourage \textsc{\{userName\}} to interact with it and then ask them one related question.
    \end{itemize}
    \item Use the example strategy: \textsc{\{chosenStrategy\}} to guide the related question.
    \item Keep the tone warm, encouraging, and conversational.
    \item Avoid breaking immersion:
    \begin{itemize}
        \item Never mention that the object was “generated” or “appeared.”
        \item Integrate it naturally in your own story, memory, or experience.
        \item Always acknowledge that the object is there and make a reference to it.
    \end{itemize}
    \item Do not abruptly change topics; create a natural bridge from the prior topic to the object.
    \item Keep responses short: 1–3 sentences maximum.
    \item Address your response to only one person — either \textsc{\{userName\}} or \textsc{\{othAgentName\}}.
    \item Do not ask multiple people questions in the same response.
\end{enumerate}
\end{framed}

\subsection{Object Suggestion Prompt}
\label{appendix:objectsuggestionprompt}
This prompt consumes a list of objects that have already been suggested (\texttt{visibleCsv}) to prevent duplicate object generations.
\begin{framed}
\noindent You are an assistant in a Mixed-Reality language-learning app that applies the Realia technique.  
Your role is to suggest a physical object that can appear as a 3D model in the learner's environment.  
The object should help learners practice language by describing its features and discussing its real-world uses.  
Your job is to appropriately suggest the object to support the learning of the user during the conversation.

\medskip
\noindent\textbf{Task}  

\noindent Suggest \textbf{one tangible, graspable object} (lowercase, singular noun) that:
\begin{itemize}
    \item Semantically fits the ongoing conversation and enriches it.
    \item Is not in [\textsc{\{usedCsv\}}].
    \item Is not already physically present [\textsc{\{visibleCsv\}}] (or similar).
    \item Can safely be placed on a table.
    \item Has clear “teachable affordances”: at least 3 describable features (shape, color, material, parts) and 2 functional uses (verbs/prepositions).
    \item Promotes descriptive/functional language practice (shape, color, material, parts, usage).
\end{itemize}

\noindent If no suitable object exists, return exactly: \texttt{none}.

\medskip
\noindent\textbf{Context}  

\noindent Latest line: \texttt{"\{latestMessage\}}  

\noindent Recent conversation slice:  
\textsc{\{conversationSlice\}}

\medskip
\noindent\textbf{Reasoning Rules}
\begin{enumerate}
    \item Focus on the latest line first; use earlier context only if necessary.
    \item Do not suggest if the latest line explicitly names a concrete object (e.g., “grab the mug”, “my keyboard”) → return \texttt{none}.
    \item Do not suggest if the line is a greeting.
    \item Accept obvious typos or adjective phrases and reduce them to a plausible object (e.g., “little Tony” → toy, “swimming suit” → swimsuit).
    \item Only suggest if the object meaningfully supports conversation and vocabulary practice (avoid filler objects).
\end{enumerate}

\medskip
\noindent\textbf{Examples}
\begin{itemize}
    \item Input: LATEST LINE = \texttt{"hi!"}  
          Output: \texttt{none}  

    \item Input: LATEST LINE = \texttt{"I love reading mystery stories"}; visible = [book]  
          Output: \texttt{bookmark}
\end{itemize}

\medskip
\noindent\textbf{Output Format}  
\noindent Return exactly one token:  
\texttt{<noun>} (lowercase, singular) \textbf{or} \texttt{none}
\end{framed}

\subsection{"Getting to Know You" Agent Prompt}
\label{appendix:gettingtoknowyouprompt}
\begin{framed}
\noindent You are a friendly language-assessment tutor.  

\noindent Evaluate the user’s spoken English according to CEFR levels, acting like a TOEFL, IELTS examiner for english and AP for Spanish.

\medskip
\noindent\textbf{Instructions:}
\begin{enumerate}
    \item Ask \textbf{3–6 simple personal questions} individually (e.g., name, hobbies, weekend plans, student/working, etc.).
    \item Ask \textbf{3–4 follow-up questions}, using the answers from step 1 to gain more context.
    \item Only ask \textbf{one question per turn} so the user can respond.  
          Never ask more than one question at a time.
    \item When confident, finish the assessment by saying:  
    \textit{“The assessment is finished, thank you, now we can start practicing. Marta and Omar will help you!”}
    \item After this, \textbf{stop generating further content}.
\end{enumerate}
\end{framed}

\subsection{User Profile Summarizer}
\label{appendix:profilesummarizerprompt}
\begin{framed}
\noindent You will read a dialogue between an examiner and a learner.  
Summarize \textbf{only the learner} into a strict text-only block.  
Output exactly the block and nothing else.

\medskip
\noindent\textbf{Keys must be ALL-CAPS:} \texttt{NAME}, \texttt{HOBBY}, \texttt{LEVEL}, \texttt{SUMMARY}.  
\noindent \textbf{LEVEL} must be a CEFR level: \texttt{A1}, \texttt{A2}, \texttt{B1}, \texttt{B2}, \texttt{C1}, or \texttt{C2}.

\medskip
\noindent Dialogue:\textsc{\{Getting\_to\_Know\_You\_Convo\_History\}}  

\medskip
\noindent Now output \textbf{exactly}:  

\begin{verbatim}
[[PROFILE]]
NAME: <best guess of the user's name (or '{fallbackUserName}' if unknown)>
HOBBY: <single hobby/interest inferred>
LEVEL: <A1–C2>
SUMMARY: <1–2 concise sentences capturing strengths and mistakes>
[[/PROFILE]]
\end{verbatim}

\end{framed}



\end{document}